\documentclass[a4paper,fleqn,11pt]{article}
\pdfoutput=1
\usepackage{amsmath}

\usepackage{amssymb}
\usepackage{array}
\usepackage{calc}
\usepackage{longtable}
\usepackage{multirow}
\usepackage{graphicx}
\usepackage{xspace}

\usepackage{mciteplus}
\usepackage[a4paper,pdfborder={0 0 0}]{hyperref}
\usepackage[format=hang,labelfont=bf,hypcap=true]{caption}
\usepackage{subcaption}
\usepackage{sectsty}
\usepackage{relsize}
\allsectionsfont{\sffamily}
\subsubsectionfont{\mdseries\itshape\large}
\makeatletter
\DeclareRobustCommand*{\bfseries}{%
  \not@math@alphabet\bfseries\mathbf
  \fontseries\bfdefault\selectfont
  \boldmath
}
\makeatother
\let\spreprint\empty
\newcommand{\preprint}[1]{\def\spreprint{\protect#1}}
\let\sinstitute\empty
\newcommand{\institute}[1]{\def\sinstitute{\protect#1}}
\makeatletter
\renewcommand{\maketitle}{\begingroup
  \null\thispagestyle{empty}%
    \ifx\spreprint\empty
      \vskip 5ex
    \else
      \flushright\large\spreprint\vskip 2ex
    \fi
    \vskip 5ex
    \flushleft
      {\sffamily\bfseries\huge\@title}\vskip 6ex
      \@author\vskip 2ex
      \ifx\sinstitute\empty
      \else
        {\small\sinstitute}
      \fi
    \vskip 3ex
  \endgroup
}
\makeatother
\renewenvironment{abstract}{\begin{center}
  {\large\sffamily\bfseries Abstract: }
  \begin{minipage}[t]{0.75\textwidth}
}{\end{minipage}\end{center}\vskip 10ex}

\numberwithin{equation}{section}

\newcommand{\Pythia}{P\protect\scalebox{0.8}{YTHIA}\xspace}

\newcommand{\Sherpa}{S\protect\scalebox{0.8}{HERPA}\xspace}

\long\def\symbolfootnote[#1]#2{\begingroup%
\def\thefootnote{\fnsymbol{footnote}}\footnote[#1]{#2}\endgroup}

\renewcommand{\c}{{\mathrm c}}

\newcommand{\e}{{\mathrm e}}

\newcommand{\g}{{\mathrm g}}

\newcommand{\p}{{\mathrm p}}

\newcommand{\s}{{\mathrm s}}

\renewcommand{\u}{{\mathrm u}}
\newcommand{\A}{{\mathrm A}}

\newcommand{\W}{{\mathrm W}}
\newcommand{\V}{{\mathrm V}}
\newcommand{\Z}{{\mathrm Z}}

\newcommand{\cbar}{\overline{\mathrm c}}

\newcommand{\sbar}{\overline{\mathrm s}}

\newcommand{\ubar}{\overline{\mathrm u}}

\newcommand{\as}{\alpha_{\mathrm{s}}}
\newcommand{\aem}{\alpha_{\mathrm{em}}}
\newcommand{\aw}{\alpha_{\mathrm{w}}}

\newcommand{\tms}{t_{\textnormal{\smaller{\smaller{MS}}}}}

\newcommand{\figRef}[1]{fig.~\ref{#1}\xspace}

\usepackage{mathtools}
\hypersetup{
  pdfauthor={Jesper Roy Christiansen,Stefan Prestel}
  pdftitle={Merging weak and QCD showers with matrix elements}
}
\author{Jesper Roy Christiansen$^1$, Stefan Prestel$^2$}
\title{Merging weak and QCD showers \\with matrix elements}
\institute{
  $^1$ Department of Astronomy and Theoretical Physics, Lund University,
  SE-22362 Lund, Sweden\\
  $^2$ SLAC National Accelerator Laboratory, 
  Menlo Park, CA 94025, USA}
\preprint{SLAC-PUB-16409\\LU TP 15-41\\MCNET-15-28}
\begin{document}
\maketitle
\begin{abstract}
  We present a consistent way of combining associated weak boson radiation in
  hard dijet events with hard QCD radiation in Drell-Yan-like scatterings.
  This integrates multiple tree-level calculations with vastly different
  cross sections, QCD- and electroweak parton shower resummation into a single
  framework. The new merging strategy
  is implemented in the \Pythia event generator and predictions are confronted
  with LHC data. Improvements over the previous strategy are
  observed. Results of the new electroweak-improved merging at a future 100 TeV
  proton collider are also investigated.
\end{abstract}
\section{Introduction}

With the Large Hadron Collider entering its 13 TeV run phase, new phenomena will be
investigated in previously inaccessible regions of phase space. Accurate calculations
for background processes in the Standard Model (SM) thus have to be reliable when 
singling out phase space regions by applying intricate analysis techniques to the 
collider data. General Purpose Event Generators~\cite{Buckley:2011ms} that are combined with multi-parton
fixed-order cross section calculations provide the most flexible assessments of SM
backgrounds. The problems to address in these methods are to
ensure that no momentum configurations are over- or under-counted, and that the perturbative
accuracy of the fixed-order matrix element calculation (ME) and parton shower (PS) resummation merge without
either being undermined. These obstacles were tackled in matching~\cite{
  Frixione:2002ik,*Nason:2004rx,*Frixione:2007vw,*Frixione:2010ra,*Torrielli:2010aw,
  *Alioli:2010xd,*Hoeche:2010pf,*Hoeche:2012ft,*Platzer:2011bc,*Jadach:2015mza,
  *Hoeche:2011fd,Alwall:2014hca} and
merging~\cite{Catani:2001cc,*Lonnblad:2001iq,*Mangano:2001xp,*Alwall:2007fs,
  *Lavesson:2007uu,*Hamilton:2009ne,*Hamilton:2010wh,*Hoeche:2010kg,*Platzer:2012bs,Lonnblad:2011xx,Lonnblad:2012ng}
methods, with next-to-next-to-leading order matching 
\cite{Hamilton:2013fea,*Karlberg:2014qua,*Hamilton:2015nsa,*Hoche:2014dla,*Hoeche:2014aia,*Alioli:2015toa} and next-to-leading order
merging~\cite{Lavesson:2008ah,*Gehrmann:2012yg,*Hoeche:2012yf,*Frederix:2012ps,*Alioli:2012fc,Lonnblad:2012ix}
currently providing the most 
precise predictions. 

It is crucial to note that these state-of-the art methods
inherit both strengths and weaknesses from less precise methods, in particular from
choices made in leading-order merging. These choices stem from uncontrolled or missing
ingredients in the parton shower. More comprehensive parton showers will lead to less
freedom and more precise predictions. This is also true for electroweak shower
resummation~\cite{Christiansen:2014kba,Krauss:2014yaa}, which are shown to be important for the accurate 
modelling of jets at large transverse momenta. In this article, we discuss how to
combine multi-jet calculations with QCD and weak parton showers in the
context of W-boson production, which highlights that

\begin{itemize}
\item[a)] processes that are disjoint at lowest-order need to be combined,
yielding a ``merging of mergings", (e.g. Drell-Yan W-boson production and QCD
$2\rightarrow2$ production both contribute to $\p \p \rightarrow \mathrm{jjW}$), 
\item[b)] weak parton showers are necessary to describe weak bosons close to or
inside jets, and to disentangle how a ``merging of mergings" should proceed,
\item[c)] merging is necessary for an inclusive prediction, and to set
starting conditions for the weak showers.
\end{itemize}
Note that if these points are not satisfactorily answered within a leading-order
merging method, then the uncertainty due to the resulting choices can only partially
be remedied by a more precise (e.g. NLO) merging method. Thus, to start with
the simplest merging approach, we improve the CKKW-L leading-order merging prescription~\cite{Lonnblad:2011xx} in the  
\Pythia~8 event generator~\cite{Sjostrand:2014zea} to address these issues. The improvements should
then carry over when merging NLO calculations. We present results for both LHC and
at a potential future 100 TeV proton collider.

In section~\ref{sec:ps_formalism} we review the weak parton showers in \Pythia, 
followed by a brief introduction to CKKW-L merging in section~\ref{sec:ckkwl_formalism}.
In these sections, we also highlight choices that have been made in both approaches. A
merging of QCD and weak showers with multi-parton cross sections, which resolves these
choices, is presented in section~\ref{sec:weak_merging}. Validations of the
implementation are presented in section \ref{sec:validation}. We then move on to discuss 
results for LHC and a future 100 TeV collider in section~\ref{sec:results} and conclude
in section~\ref{sec:conclusions}.

\section{Weak parton-shower formalism}
\label{sec:ps_formalism}

Scattering processes containing massless partons with very different transverse
momenta exhibit logarithmic divergences that limit the applicability of perturbative
calculations. Fortunately, it is possible to derive factorisation 
theorems and sum logarithmic terms to all orders in perturbation theory. 
This leads to a reliable, finite calculation with an extended range of 
validity. Leading-logarithmic contributions can be summed in a process- and
observable-independent fashion by using PS programs\footnote{Note 
that many other universal subleading effects are also included in parton
showers, and that for specific observables, better accuracy, than the formal
leading log, can be 
achieved~\cite{Catani:1990rr,*Nagy:2009vg}.}.

Large scale hierarchies involving massless particles still lead
to logarithmic enhancements that should, for a stable prediction, be summed 
to all orders in perturbation theory. The resummation of logarithmic 
electro-weak enhancements becomes important when processes contain low 
transverse-momentum weak bosons and jets with transverse momentum much larger
than the boson mass.
It has been shown in fixed-order calculations
that weak Sudakov corrections can 
indeed become relevant at LHC
energies~\cite{Dittmaier:2012kx,*Ciafaloni:2006qu,*Moretti:2005ut,*Denner:2001mn,*Baur:2006sn,*Campbell:2015vua}
and especially when considering potential future 
proton colliders~\cite{Mishra:2013una}.
Including all-order electroweak effects in flexible,
commonly used programs facilitates realistic studies of these effects. 

General Purpose Event generators include an approximation of 
all-order effects with the help of parton showers.
Parton showers produce all-order (QCD or QED) results by resumming real-emission
corrections into exponentiated no-emission probabilities. These no-emission
probabilities are related to Sudakov form factors by application of DGLAP
evolution~\cite{Gribov:1972ri,*Dokshitzer:1977sg,*Altarelli:1977zs}. Electro-weak resummation is a natural extension to the QCD and QED
showering. EW showers have, due to the dominance of QCD effects, only recently
been investigated in event
generators~\cite{Christiansen:2014kba,Krauss:2014yaa}. The EW shower allows
for an equal treatment of QCD, QED and weak radiation and naturally 
includes competition between emissions of gluons, photons or weak gauge
bosons. 
In this section a short summary of the major
issues are given, with a specific focus on aspects relevant for merging
parton showers with matrix element calculations.  

There are two major differences between $\gamma$ emission and $\W^\pm$
emissions. Firstly, the emission of a $\W^\pm$ boson changes the flavour of the
radiator, and secondly, the $\W^\pm$ is massive. Flavour changes are handled
according to the CKM matrix, with additional care needed for the
evaluation of PDFs. A phase space mapping for emissions of massive
particles was previously given in the context of a Hidden-Valley
PS model~\cite{Carloni:2011kk}, and the weak showers can directly
reuse the corresponding structures in \Pythia.

The massive phase space does not include the collinear and soft divergences, 
since the weak boson has to carry at least enough energy to be on its mass 
shell. The introduction of mass should also affect the PS splitting
kernels. The normal massless collinear approximation in the PS is therefore 
not sensible for radiation of weak gauge bosons. Thus, a complicated assessment 
of mass effects seems necessary. However, this can be avoided if the full, 
massive matrix elements are used as splitting kernels. The weak parton
shower in \Pythia thus heavily relies on ME corrections~\cite{Bengtsson:1986hr,*Seymour:1994df,*Miu:1998ju,*Norrbin:2000uu}. 
All emissions are corrected with a fully massive $2\rightarrow 3$ matrix element.
The corrections vary depending on the type of process -- an $s$-channel
process will for example carry a different correction factor than a $t$-channel
process. Different corrections are mandatory in order to
obtain a reasonable agreement between the PS prediction and gauge-invariant 
subsets of the full ME
calculations (including all interferences). Note however that the weak parton
shower only recovers the  $\p\p\rightarrow \mathrm{jjV}$ matrix elements
(where $\V=\W,\Z$). The weak shower further omits interference terms between
different fermion lines. It is further only possible to choose the correct ME
correction if the underlying type of process is known. Therefore the weak
shower resorts to (artificial) choices if the evolution is not started from a
$2\rightarrow1$ or a $2\rightarrow 2$ process. As will be described in detail
later, this problem is resolved through the introduction of PS histories.

The introduction of weak parton showers leads to potential double counting
in an inclusive event generation. If the desired process is 
dijet + $\W^\pm$ it can be interpreted in two ways: either as a Drell-Yan-like 
$\W^\pm$-boson process followed by two QCD emissions, or as a $2
\rightarrow 2$ QCD process radiating a $\W^\pm$-boson. Allowing these two
possibilities to separately cover the full phase space results in double 
counting. Disallowing QCD emissions above the weak boson mass for
Drell-Yan-like processes would ameliorate this double counting, yet result in 
an unconvincing data description of the pure PS result. Instead,
a strategy using the $k_\perp$ jet algorithm was adopted. If the jet separation
between a $\W^\pm$-boson and a parton proves the minimal scale, then the
events are removed from the Drell-Yan-like sample. Conversely, dijet
states whose minimal jet separation is between partons are removed from
the $2\rightarrow 2$ QCD event sample. This artificial separation will be
corrected upon merging weak and QCD showers with multi-parton matrix elements.

The weak coupling to a fermion depends on its spin. In the weak shower this
is handled in a simplistic way, by assigning each fermion line a randomly
chosen spin. The spin is then kept fixed through the whole PS. For a single
weak emission it corresponds to using averaged spin results, but it introduces
a slight enhancement for multiple weak gauge bosons to be emitted from the
same fermion line.

The overall performance of the weak PS is surprisingly good. It is
capable of describing a large number of measurements that earlier has only
been possible to describe with merged samples. Notable the rate of $\W+n$ jets
could be described up to the highest measured ($n=7$) bin without using more
than $\W+2$ jet matrix elements. But the PS still does not
provide a perfect description, and especially the description of angular 
distributions (e.g. $\Delta \phi$ between leading and second leading jet) is
poor. Merging is expected to significantly improve the results.

\section{Merging}
\label{sec:ckkwl_formalism}

Many interesting multi-jet observables at a hadron collider are difficult to predict with
calculations containing a fixed (or limited) number of outgoing partons in fixed-order
perturbation theory. Parton showers are then necessary to spread the fixed-order 
calculation over a broader multi-jet phase space. Standard examples are jet
rates, where fixed-order calculations become prohibitively expensive, or azimuthal separations
between (reconstructed) heavy bosons and a hardest jet, which are naturally sensitive to
momentum configurations with a variable number of hard jets~\cite{Wobisch:2015jea}. To describe
such genuine multi-jet observables, many multi-parton calculations need to be combined
into an inclusive sample describing configurations with $n \leq N$ jets with fixed-order
accuracy, where $N$ should be as large as possible.

\begin{figure}[th]
  \centering
  \includegraphics[width=0.4\textwidth]{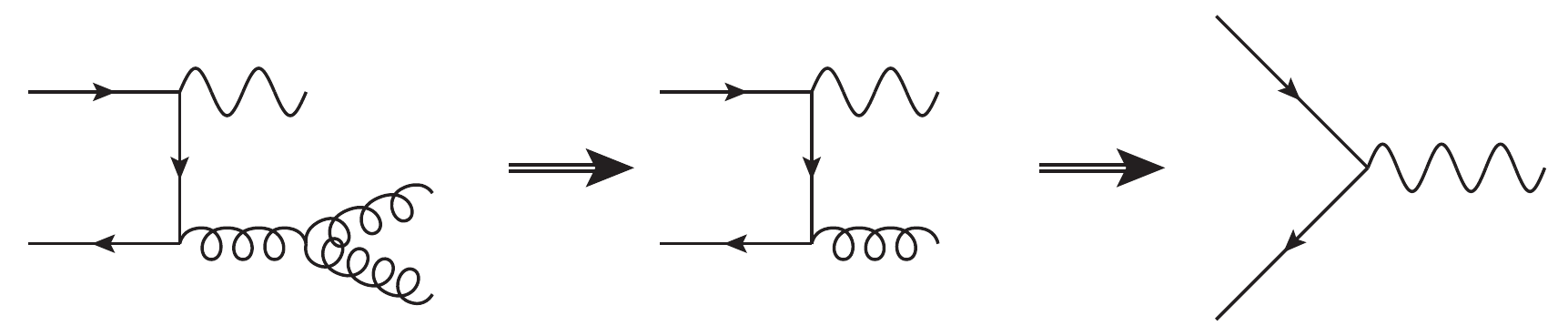}
  \hspace{1.5cm}
  \includegraphics[width=0.4\textwidth]{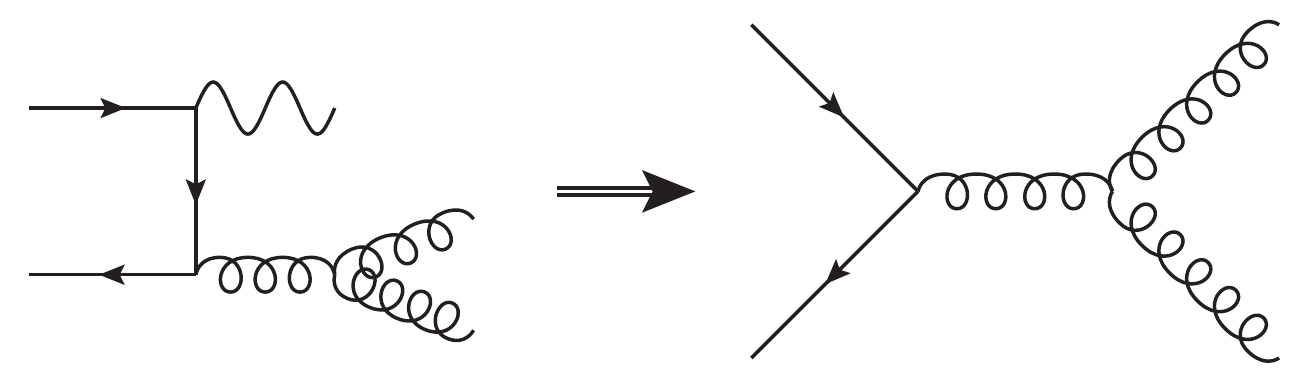}
  \caption{\label{fig:histories} Two examples of possible histories for a $\p
    \p \rightarrow \g \g \W$ process. The two histories have different hard
    processes, either as a Drell-Yan process (left) or as a $2\rightarrow2$ QCD
  process (right).}
\end{figure}

Matrix element merging is a process-independent method that invokes the PS to
facilitate this combination. The main steps in a merging procedure 
are\footnote{We will call a physical flavour, colour and momentum configuration a "state".}:
\begin{itemize}
\item choose a "history" of intermediate states through which a pre-calculated input
multi-parton state has evolved from a lowest-multiplicity state 
(see e.g.\ \figRef{fig:histories}),
\item use this history to make the state exclusive (i.e.\ additive) by calculating
and applying the necessary no-emission probabilities (which are intimately linked to
Sudakov form factors),
\item reweight the input state with additional factors (e.g. $\as(p_\perp^2)$, PDF weights) that would have been applied by
the PS, had it produced the input state by following the history of intermediate
states (this is necessary to not impair the accuracy of the PS, or the event 
generator prediction more generally),
\item combine the result of all such post-processed input states for all parton multiplicities.
\end{itemize}
This immediately highlights that omissions in the shower lead to uncertainties
in the merging prescription, which are commonly disposed of by judicious selection. 
Since we are interested in combining with weak parton showers, let us look at producing an
inclusive sample of W-boson + $N$ jets through CKKW-L merging, and assume $N\leq 3$ for
simplicity.

In this example, the lowest-multiplicity state (W-boson production) should be used to
describe very inclusive observables like e.g.\ the W-boson rapidity. The interface to the
PS is straightforward since no partons are present initially -- we only have to ensure that
the PS does not produce hard jets, as such configurations should be covered
by higher-multiplicity matrix elements. This leads to the introduction of a ``merging
scale" with arbitrary functional definition and value $\tms$. States that are classified as
``below" $\tms$ will be produced by showering, while states ``above" $\tms$ are governed
by higher-order matrix elements. Any functional form of the merging scale should be allowed, as long
as the function acts as a regularising cut on the fixed-order input calculations.
Commonly used merging scale definitions are the minimum of all jet separations in the
$k_T$ algorithm~\cite{Catani:1993hr}, or the minimum of parton
shower evolution variables measured on the state. Merging methods have to ensure that
the dependence of exclusive and inclusive observables on the merging scale are small. For 
inclusive jet observables, the merging scale dependence can be removed to
reasonable accuracy\footnote{The method of section~\ref{sec:weak_merging}
  will, when applied in 
unitarised merging~\cite{Lonnblad:2012ng}, allow to cancel the $\tms$ dependence of inclusive
cross sections exactly.}.

Coming back to our example, the next calculation
to be added is W-boson in association with one parton. As outlined above, a PS
history has to be chosen for such states. These histories are well-defined if the QCD parton
shower can (at least in principle) cover the full single-emission phase space. In order to 
pick all histories in the proportion in which the PS would have produced the
output state, the probability for a specific history is given by the product
of splitting functions characterising each intermediate evolution step. This
reduces the merging scale 
dependence of exclusive observables. Upon choosing a history, it is simple to reweight
with no-QCD-emission probabilities (i.e.\ QCD Sudakov factors) and to account for the
dynamic renormalisation and factorisation scales used in the PS evolution.
Using the shower directly to produce the no-QCD-emission probabilities reduces the
$\tms$ dependence. The starting conditions for PS emissions off the W + parton state
are uniquely determined by the chosen history. 

Including a W-boson in association with two partons uncovers further 
uncertainties, because no ordered PS will cover the full double-emission phase
space. Thus, some states accessible to the fixed-order calculation
will not yield any ordered PS history\footnote{In the following, we will use the phrase ``unordered states"
when talking about input states that do not yield any ordered PS history.}. The reweighting of such a
state is ambiguous due to ambiguous renormalisation and 
factorisation scale choices. Although this ambiguity has very small numerical impact
for inclusive observables, it can have an uncomfortably large impact on more exclusive
observables~\cite{Lonnblad:2011xx}. Furthermore, some flavour configurations are inaccessible
to a QCD parton shower, meaning that no PS history can be reconstructed. Ambiguities
in the treatment of such genuine non-shower (commonly called incomplete) states have
vanishingly small impact on inclusive observables and yield only very minor variations
of exclusive observables~\cite{Lonnblad:2011xx}. A precise method should however avoid having
to make choices. The PS starting conditions are fixed once a history is
chosen.

No new problems occur for higher-multiplicity processes. The issues related to
unordered states outlined in the last paragraph can be aggravated in
more exclusive observables, however, as the PS phase space coverage will be worse
for higher multiplicities. It is still important to remember that merging offers
a consistent way to set the PS starting conditions for multi-parton states --
which is not the case in plain (QCD or EW) parton showers. 

\section{Weak showers and the merging of merged calculations}
\label{sec:weak_merging}

In the previous sections, we have seen that the construction of weak parton showers
as well as multi-jet merging involves compromises. Summarising the
most severe choices, we have addressed that:
\begin{itemize}
\item Weak showers are currently limited to dijet processes, while inclusive
      predictions require an ambiguous mixing with Drell-Yan-like configurations.
\item Matrix element merging is ambiguous starting at W-boson + two partons,
      leading to uncomfortable compromises related to unordered states and incomplete
      histories.
\end{itemize}
The combination of weak parton showers and multi-jet merging remedies these 
deficiencies and should provide a more physical picture of multi-parton
states. This will further mean that matrix element merging, which is
usually regarded to realise corrections to one underlying process, is
generalised to incorporate many underlying processes that mix at higher
perturbative orders.

To explain the reasoning behind our new merging scheme, let us look at states 
including one weak boson and two final state partons for 
illustration. If the outgoing partons have very different transverse momenta, 
and if the $p_\perp$ of the W-boson is thus comparable to the
$p_\perp$ of the harder parton, then it is natural to associate the partons
with DGLAP evolution of the incoming beams. For a reliable 
perturbative prediction, a fixed-order calculation with large scale separation
should then be supplemented with no-emission probabilities resumming unresolved
QCD emissions. This W-boson + two parton state is an example for corrections
to W-boson production.

If the state instead contains two partons with comparable
and large $p_\perp$ and a W-boson with small transverse momentum, it
is prudent to resum large logarithms associated with the difference between
the parton and W-boson transverse momenta. Then Sudakov form factors resumming 
the dominant weak virtual corrections need to be supplemented. Hence, the
two-parton + W-boson state is an example for corrections to dijet
production. This correction can be approximated by weak parton showers.

Thus we find that, when going to $\mathcal{O}\left(\as^2\aw\right)$, it is not 
possible to disentangle QCD corrections to W-boson production and weak corrections
to dijet production\footnote{$\aw$ is used as proxy of the weak coupling constant, which will differ depending on
the type of the radiated boson.}. Beyond $\mathcal{O}\left(\as^2\aw\right)$, W-boson
production and dijet production share a single evolution, so that only a 
combined treatment of these two processes (which are disjoint at lowest 
multiplicity) will yield a satisfactory prediction. This means that  
it is necessary to combine multi-jet merged corrections to W-boson production
with multi-jet merged corrections for dijet production. These corrections
then mix by virtue of weak showers\footnote{To be fully consistent, it would 
be necessary to be inclusive both in QCD and EW outgoing particles. A 
complete description of a $\A,\mathrm{B} \rightarrow 4$ particle state
should contain any admixture of W-bosons and partons with four or less outgoing
particles. This article only addresses the combination of dijet and W-boson 
production, since processes with multiple W-bosons are rare, and assuming that 
many radiated W-bosons escape detection further seems unrealistic.}. This
in a sense constitutes a ``merging of mergings".

Summarising, we have argued that a clean description of W + jets states
necessitates a combination of QCD no-emission probability-reweighted corrections to to W-boson
production and weak no-emission probability-reweighted dijet production. Within this framework,
it is possible to address and amend the choices in weak showering and merging
that we have previously highlighted.

The first feature of the combined merging is the possibility to recombine
W-boson radiation with other partons. As a natural consequence of this the 
lowest-multiplicity process is, as desired, no longer forced to be a 
colour-singlet Drell-Yan-like state if the input state contained W-bosons. The
new clustering is illustrated in \figRef{fig:histories}, where two 
very different possible histories are shown\footnote{Note that in the
\Sherpa~\cite{Gleisberg:2008ta} event generator, this method is also used, albeit without taking
the corresponding weak no-emission probabilities into account~\cite{Schalicke:2005nv}. Also,
histories are picked by choosing probabilistically at each history node, while \Pythia
generates all histories before choosing a whole path probabilistically.}.
The decision which of these histories to choose should again ensure that the
merging scale variation of exclusive observables is small. This means we 
should attempt to answer the question \emph{how would the (QCD+EW) parton shower
have produced this state?} The answer will minimise merging artefacts at the
boundary between PS and fixed-order ME regions. With the parton
shower probabilistically sampling all ways to evolve into a particular state, we 
again decide to pick histories with different underlying process
probabilistically. For instance the two histories shown 
in \figRef{fig:histories} would have the following probabilities:
\begin{equation}
  \mathcal{P}_\textnormal{left path} = \frac{
  \mathcal{P}_{\textnormal{\smaller QCD FSR}}^{(1)}\,
  \mathcal{P}_{\textnormal{\smaller QCD ISR}}^{(2)}\,
  \mathcal{P}_{\textnormal{weak {\smaller W} production}}^{(3)}
}{\smashoperator[r]{\sum_{\textnormal{\smaller all paths}}}~\,\, \prod_{\substack{\textnormal{\smaller nodes j}\\\textnormal{\smaller in path i}}}\mathcal{P}^{(j)}_{\textnormal{\smaller type}} }
\qquad\qquad
 \mathcal{P}_\textnormal{right path} = \frac{
\mathcal{P}_{\textnormal{weak {\smaller ISR}}}^{(1)}\,
\mathcal{P}_{\textnormal{{\smaller QCD} jet production}}^{(2)}
   }{\smashoperator[r]{\sum_{\textnormal{\smaller all paths $i$}}}~\,\, \prod_{\substack{\textnormal{\smaller nodes j}\\\textnormal{\smaller in path i}}}\mathcal{P}^{(j)}_{\textnormal{\smaller type}} }
\end{equation}
where $\mathcal{P}^{(j)}_{\textnormal{\smaller type}}$ indicates probability
associated to the $j$'th clustering in the path, with ``type" indicating what
type of transition occurred. 

The coupling between fermion and weak gauge bosons depends on the spin of the
fermion. To capture this effect in 
the merging, histories for all possible spin assignments for fermions are considered. 
One improvement of spin treatment could be to use fully spin dependent input
matrix elements. However, in order for this to be consistent, improvements in 
the spin handling within the PS would be required.

An additional constraint on the probabilities comes from insisting on 
$p_\perp$-ordered histories: clusterings of states with lower multiplicity
have to have a larger $p_\perp$ than clusterings of higher-multiplicity states. For instance, if the event consists of two
hard jets and a soft W-boson, it is very unlikely to cluster it to a
Drell-Yan hard process and obtain a $p_\perp$-ordered clustering sequence. Within a combined
merging of dijet and W-boson production, the dominant scale hierarchies are
correctly identified. Hence, the amount of unordered states is drastically reduced.

The necessity for weak clusterings and the weak showering effects introduces 
two new weights to the merging procedure: an $\aw$ weight and the weak
no-emission probability. The $\aw$ weight is required because a dynamical
scale setting is also assumed when evaluating $\aw$.

The weak no-emission probability can be generated by trial showering. To treat
QCD-like and electroweak emissions on equal footing, we include W-bosons
in the merging scale definition, meaning that "soft" W-bosons will be generated
by the PS, while "hard" W-bosons are generated with the help
of a fixed-order matrix element generator. This also means that in non-highest
multiplicity states, any first PS response producing states with
a hard W-boson (or, of course, hard QCD emissions) will lead to an event 
rejection. The impact of the weak no-emission probabilities can, due to the large 
W-boson mass and the small value of $\aw$, be minor for many observables. 
However, for observables with large hierarchies between the scales associated
to QCD emissions and scales of EW effects, larger effects are anticipated.
An idealised observable highlighting weak resummation effects would be very
inclusive over multi-parton states and fully exclusive for weak emissions 
(i.e.\ all weak bosons can be resolved). We will return to this in the
result section, where the effect of the weak no-emission probabilities at 
a future 100 TeV collider is considered.

While this new method leads to a more physical description of multi-parton
states in association with W-bosons, it should be noted that the formal
accuracy of neither QCD resummation nor fixed-order calculation is improved.
However, this merging for the first time supplements arbitrary multi-jet 
states with weak resummation effects within a matrix-element-merged prediction.
Thus, the electro-weak all-order structure improves over previous results.

To round off this section, remember that merging methods allow the combination
of different jet multiplicities. A combination is only possible because the inclusive fixed-order input
states (describing $N$ or more particles) are converted into exclusive
calculations (describing exactly $N$ resolved particles) by supplementing
no-emission probabilities which resum logarithmic enhancements due to large scale
hierarchies. It would thus at first glance seem that a state containing two soft
QCD emissions at vastly different scales and a W-boson with transverse
momentum commensurate with the larger jet scale is in some sense "more exclusive" than a dijet
state with jets of similar $p_\perp$ and a soft W-boson. In the former case
two scale differences require resummation, while in the latter, only one
hierarchy has to be considered. However, note that the dijet cross section
is not well-defined unless jet cuts are applied. These cuts make the 
cross section exclusive in the sense that at least two jets above a resolution
scale are required. That the cross section contains exactly the desired
number of jets (and no further resolved jets) is then again achieved by reweighting with no-emission probabilities.
As an aside, note that multi-parton interaction (MPI) models~\cite{Sjostrand:1987su,*Alekhin:2005dx,*Bahr:2008dy} are derived from the 
condition that the dijet cross section needs to be regularised, and that this
regularisation can be achieved in the same way that no-emission probabilities
regularise parton-shower real-emission cross sections. The
no-MPI-probabilities motivated by this argument should be correctly included in any merging scheme to
ensure that the input states do not overlap with MPI, adding yet another layer
of exclusivity. Our implementation in \Pythia includes a consistent handling
of interleaved MPI~\cite{Sjostrand:2004pf} as outlined in~\cite{Lonnblad:2011xx}.

\section{Validation}
\label{sec:validation}

When developing an improved merging scheme, detailed tests validating
the method and implementation are necessary. We have tested that the new
implementation recovers the correct scales, probabilities and underlying
states by directly comparing a reconstructed PS evolution history against the 
evolution as picked by the parton shower. Such technical comparisons are of
course not particularly enlightening for the reader, so that below, we will focus on 
two hopefully convincing tests.

\begin{figure}[tp]
  \centering
  \includegraphics[width=0.45\textwidth]{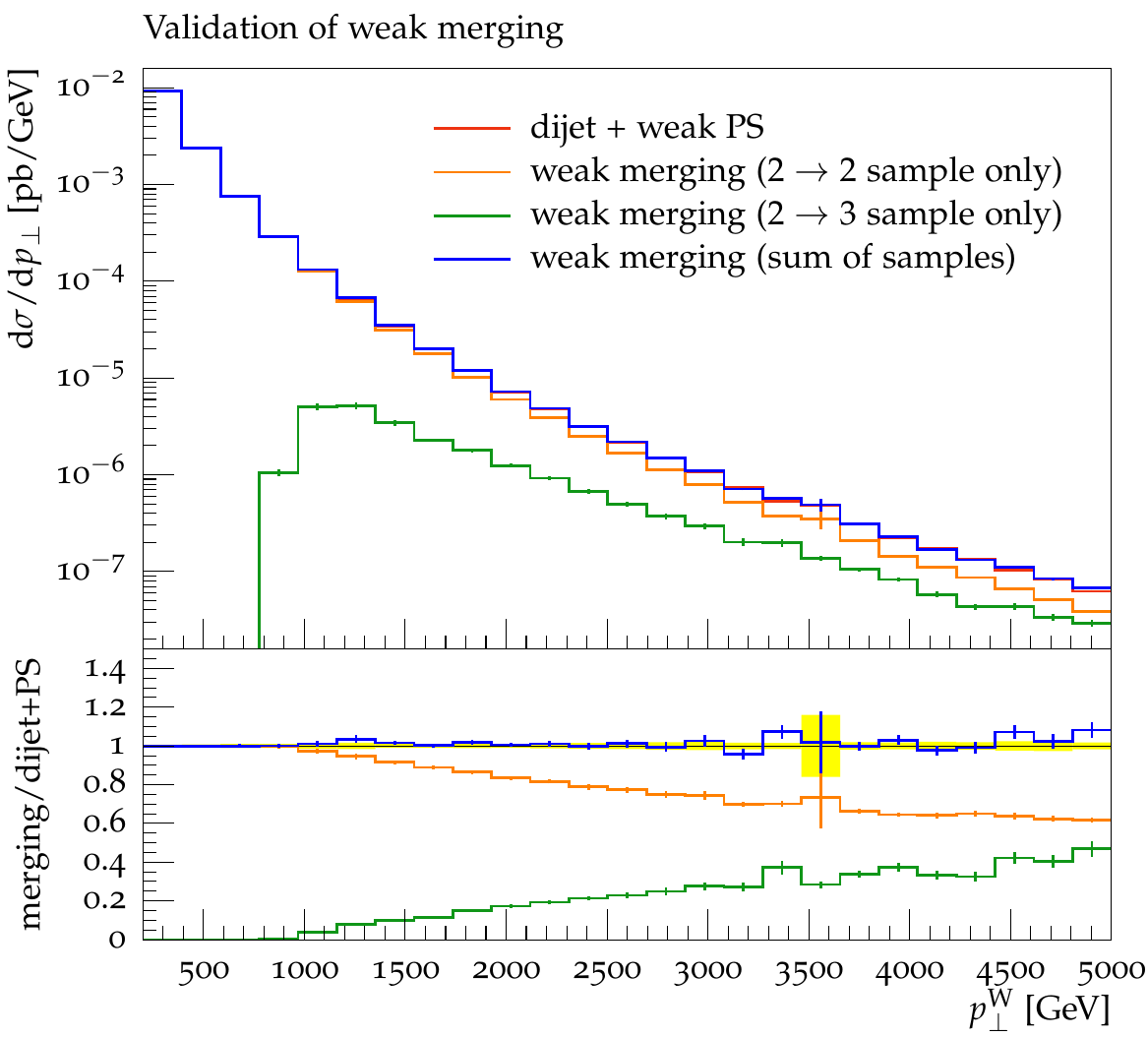}
  \includegraphics[width=0.45\textwidth]{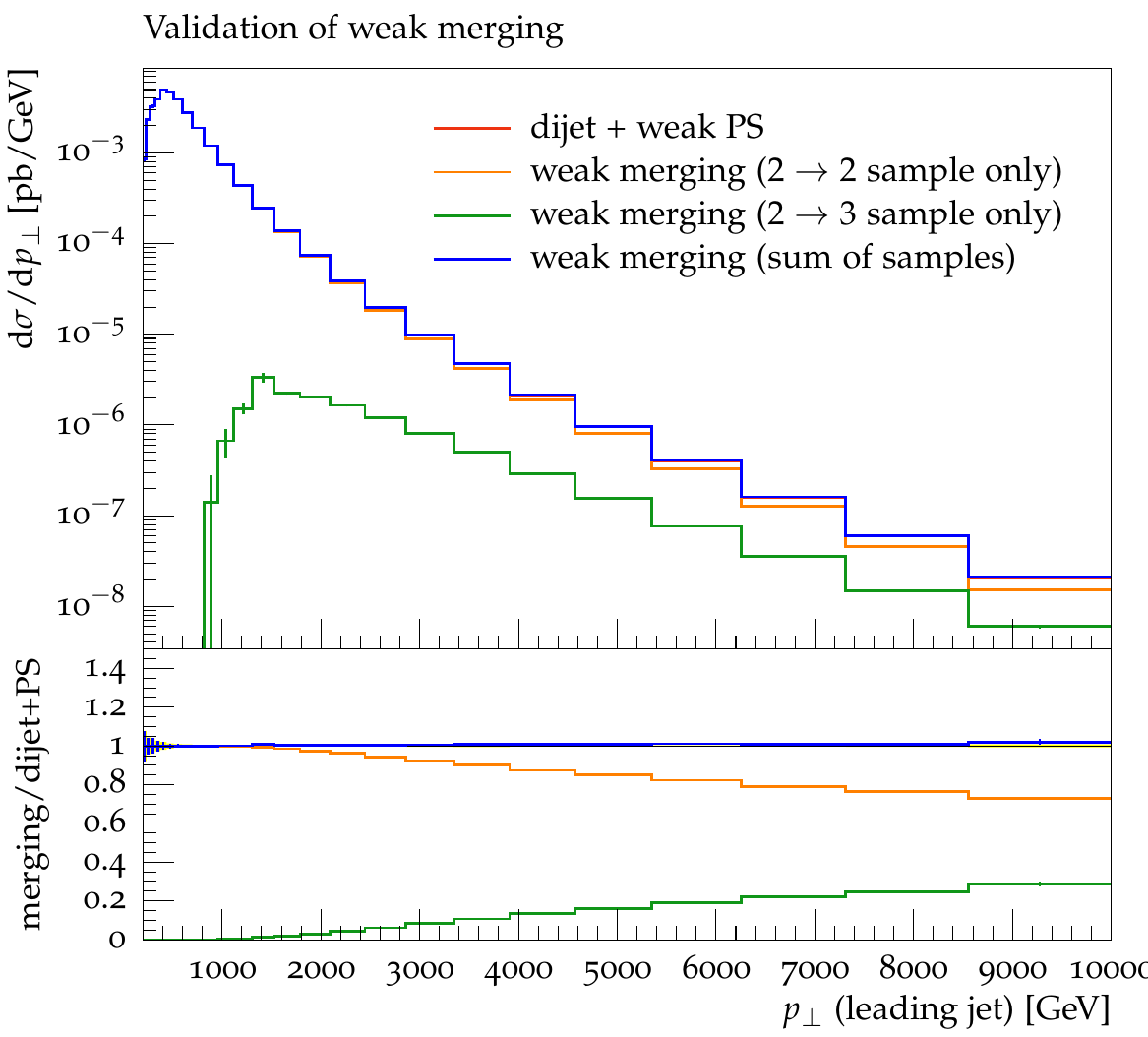}
  \caption{\label{fig:sudakovVal}The figure shows the cross section for
    $\u\ubar \rightarrow \s \cbar \W^+/\sbar \c \W^-$ as a function of
    respectively $p_\perp$ of the $\W^\pm$ (left) and $p_\perp$ of leading jet
    (right). The cross section is
    calculated in two ways: Either through merging of $2\rightarrow 2$ and
    $2\rightarrow3$ MEs, or as a $2\rightarrow2$ ME with weak shower. }
\end{figure}

The weak PS relies on ME corrections for the process $pp\rightarrow \mathrm{jj}\W$. As 
such, an excellent agreement between merged and default weak PS results for 
such $2\rightarrow3$ processes is expected. We illustrate the agreement for
the process $\u \ubar \rightarrow \s \cbar \W^+/\sbar \c \W^-$, only including the 
$\mathcal{O}\left(\alpha_s^2\aw\right)$ contributions as fixed-order inputs. This is an
$s$-channel process, where only final state radiation is possible (assuming a diagonal CKM
matrix). $\aem$ was set to 0.1 to increase the statistics and the merging scale was
1000 GeV. The merged curve and the weak PS agree nicely over the whole kinematic range, as illustrated by both the 
W-boson $p_\perp$ and the leading jet $p_\perp$ distributions
(\figRef{fig:sudakovVal}). The merged result thus correctly applies all
factors present in the weak PS resummation. 

\begin{figure}[tp]
  \centering
  \includegraphics[width=0.45\textwidth]{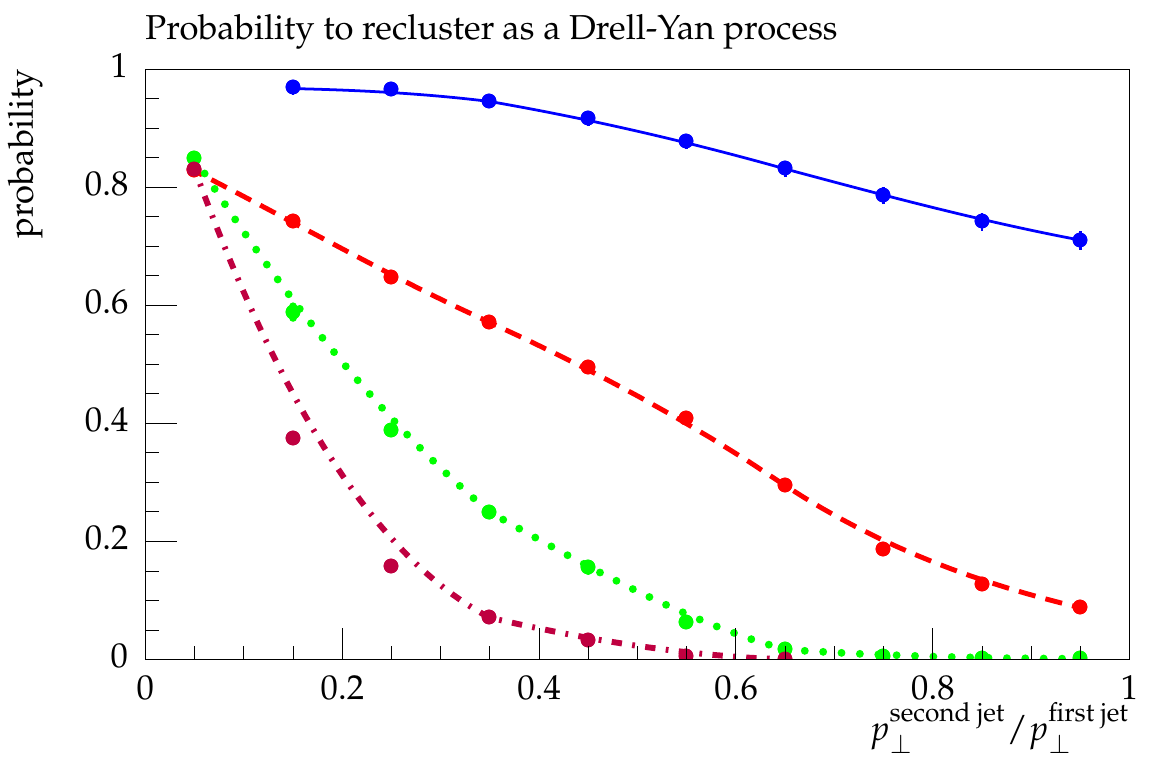}
  \includegraphics[width=0.45\textwidth]{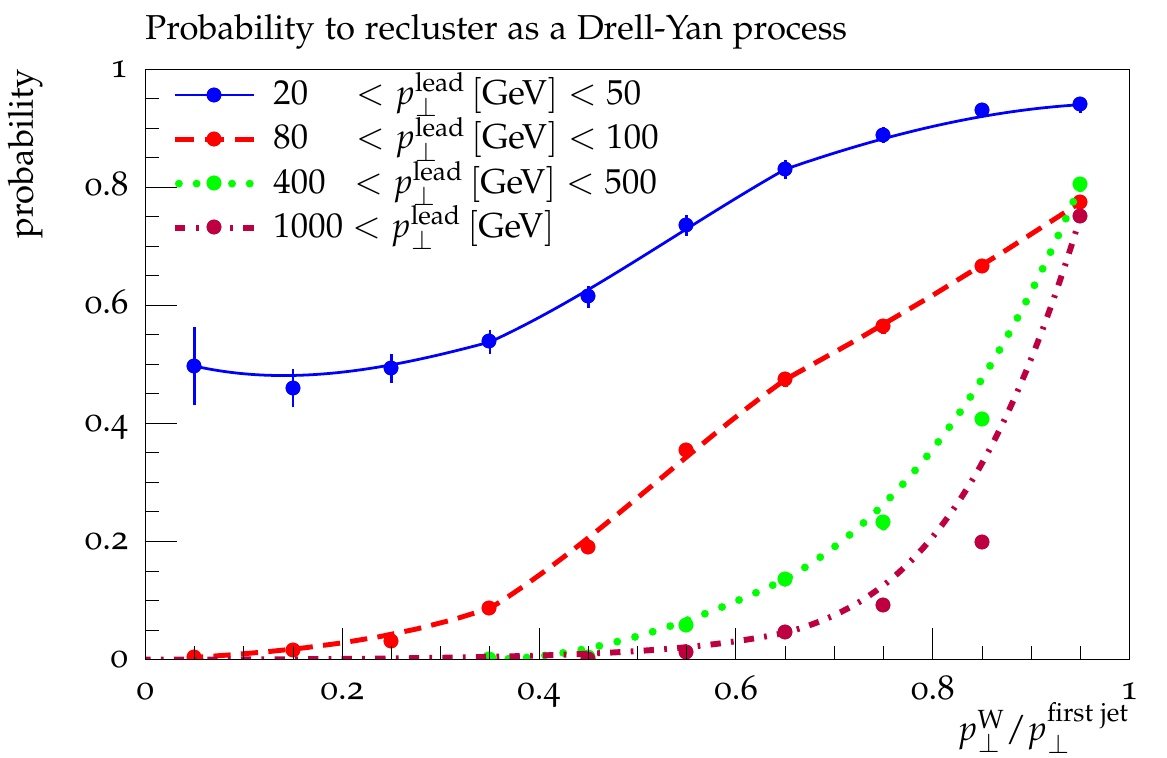}
  \caption{\label{fig:competition}The figure shows the competition for either
    clustering a $\p\p \rightarrow \mathrm{jj}\W$ process as a Drell-Yan process or
    a $2\rightarrow 2$ QCD process. The probability is shown as a function of
    the fraction between either the second leading jet $p_\perp$ divided by the leading jet
    $p_\perp$ (left) or the fraction between the W $p_\perp$ and
    the leading jet $p_\perp$ (right). The minimum $p_\perp$ for any jet is 5 GeV and
    the centre-of-mass energy is 7 TeV. The vertical lines indicate the
    statistical MC uncertainty and smooth curves have been added as a visual help. }
\end{figure}

To further validate the implementation, \figRef{fig:competition} shows the
probability with which states are identified as corrections to a Drell-Yan-like
or a $2\rightarrow 2$ QCD hard scattering. Each path is expected to dominate in a specific
region of phase-space. If the scales associated to jet production are low and
exhibit a hierarchy, then a Drell-Yan-like underlying process should be expected.
States with two hard jets at comparable scales should yield a $2\rightarrow 2$ QCD
underlying process. We investigate this expectation on the process
$\p\p \rightarrow \mathrm{jjW}$, using different $p_\perp$ cuts on the 
leading jet (fig. \ref{fig:competition}). As expected, the lower the
$p_\perp^{\text{leading jet}}$ cut is, the more likely states will lead to a
Drell-Yan-like underlying process. Conversely, for a fixed leading 
jet $p_\perp$, softer $\W$ $p_\perp$'s and more back-to-back jet systems yield
predominantly QCD $2\rightarrow 2$ scatterings as underlying process.

\section{Results}
\label{sec:results}

This section presents predictions of merging QCD+EW showers with multi-parton
matrix elements. We begin by comparing with studies from both ATLAS and CMS
and follow up by a study of the weak no-emission probability at 100 TeV.

\subsection{Comparison with LHC data}

\begin{figure}[tph]
  \centering
  \includegraphics[width=0.45\textwidth]{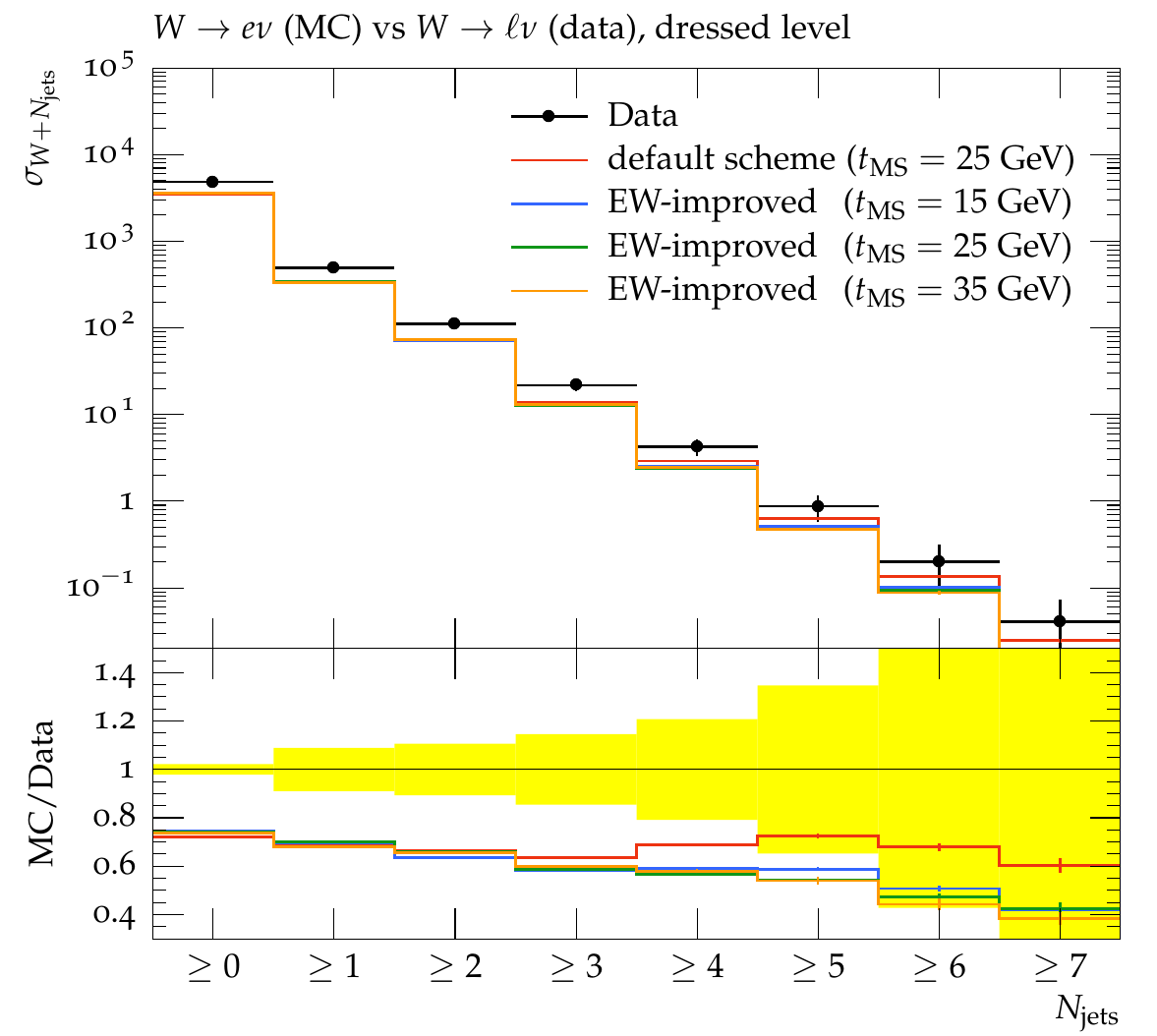}
  \includegraphics[width=0.45\textwidth]{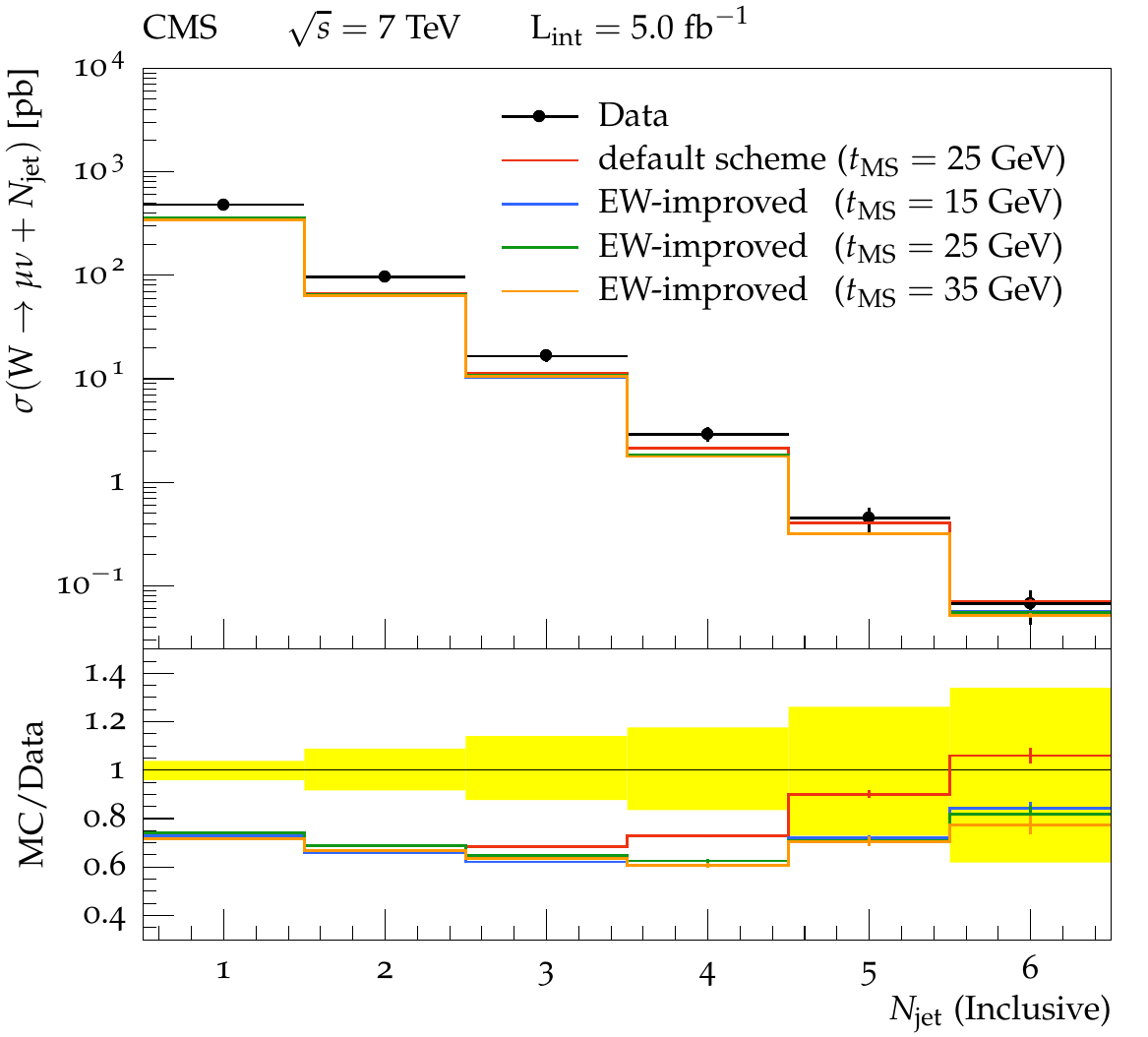}
  \caption{\Pythia predictions in comparison to ATLAS data~\cite{Aad:2014qxa}
    (left) and CMS data~\cite{Khachatryan:2014uva} (right)
    for W + jets as a function inclusive jet multiplicity. The yellow error
    band indicates the one sigma experimental uncertainty and the vertical
    line on the MC prediction is the statistical MC uncertainty.
    \label{fig:wJetMult}}
\end{figure}

In this section, we contrast results of the default CKKW-L merged prescription
in \Pythia and the new QCD+EW merging with LHC data.

To compare against LHC data, we merge five tree-level event samples for 
W-boson $+\leq 4$ jets, generated 
with MadGraph5$\_$aMC@NLO~\cite{Alwall:2014hca} using the CTEQ6m PDF 
set~\cite{Pumplin:2002vw}. The merging scale was defined
as the minimum of all \Pythia transverse momentum separations between partons, 
while no cut was applied to the the W-boson. This means that the phase
space for real weak parton showers is vanishing, thus making the inclusion of 
pure-QCD samples unnecessary\footnote{Real W-boson emissions can only enter
through 5-parton events or, for lower-multiplicity events, if a QCD emission 
below $\tms$ is followed by a weak emission. Both contributions
have a negligible effect.}. Thus, this setup can be used in particular to
check the impact of the ``weak clustering" outlined in
section~\ref{sec:weak_merging}.
The Monash tune~\cite{Skands:2014pea} was used, but with $\as(M_Z)$ lowered to
$\as(M_Z) = 0.118$.

Our results only contain tree-level normalisation, and an overall rescaling due
to virtual corrections is missing. The results are therefore not expected  to
match the normalisation of the data. We choose to not rescale our results since
we believe that presenting unnormalised experimental data adds additional 
information and should be encouraged. We do not want to undermine such efforts
by rescaling tree-level results. The differential shape of the data should 
however be described by a tree-level merged prediction (i.e.\ the ratio between
the data and the prediction should be flat for all distributions). All
the data comparisons are done using the Rivet framework~\cite{Buckley:2010ar}.

In the following, we will refer to the default CKKW-L implementation in \Pythia
as ``default scheme", while the new QCD+EW merging will be called ``EW-improved"
scheme. The EW-improved results are shown for three different merging
scales, $\tms = 15$ GeV, $\tms = 25$ GeV and $\tms = 35$ GeV. The uncertainty
due to this merging scale variation is very small for all observables we have
investigated, and is nearly indistinguishable from statistical fluctuations
for the observables below. The very small variation is a result of the PS both
correctly recovering the $\W+1$ j matrix element as well as hard (dijet-like) 
parts of the $\W+2$ j matrix elements, thus pushing the merging scale dependence
to yet higher orders.

\begin{figure}[tph]
  \centering
  \includegraphics[width=0.45\textwidth]{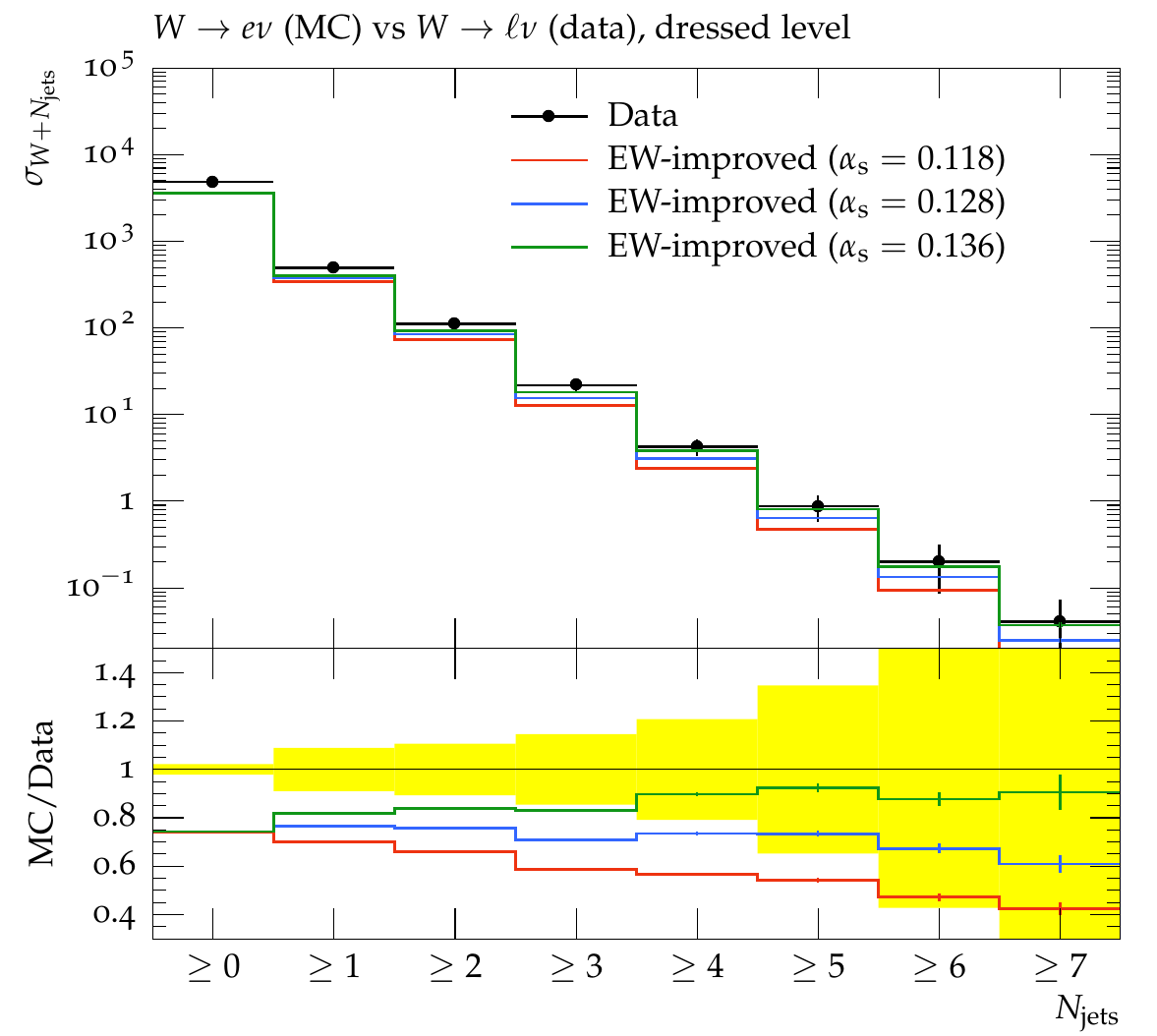}
  \includegraphics[width=0.45\textwidth]{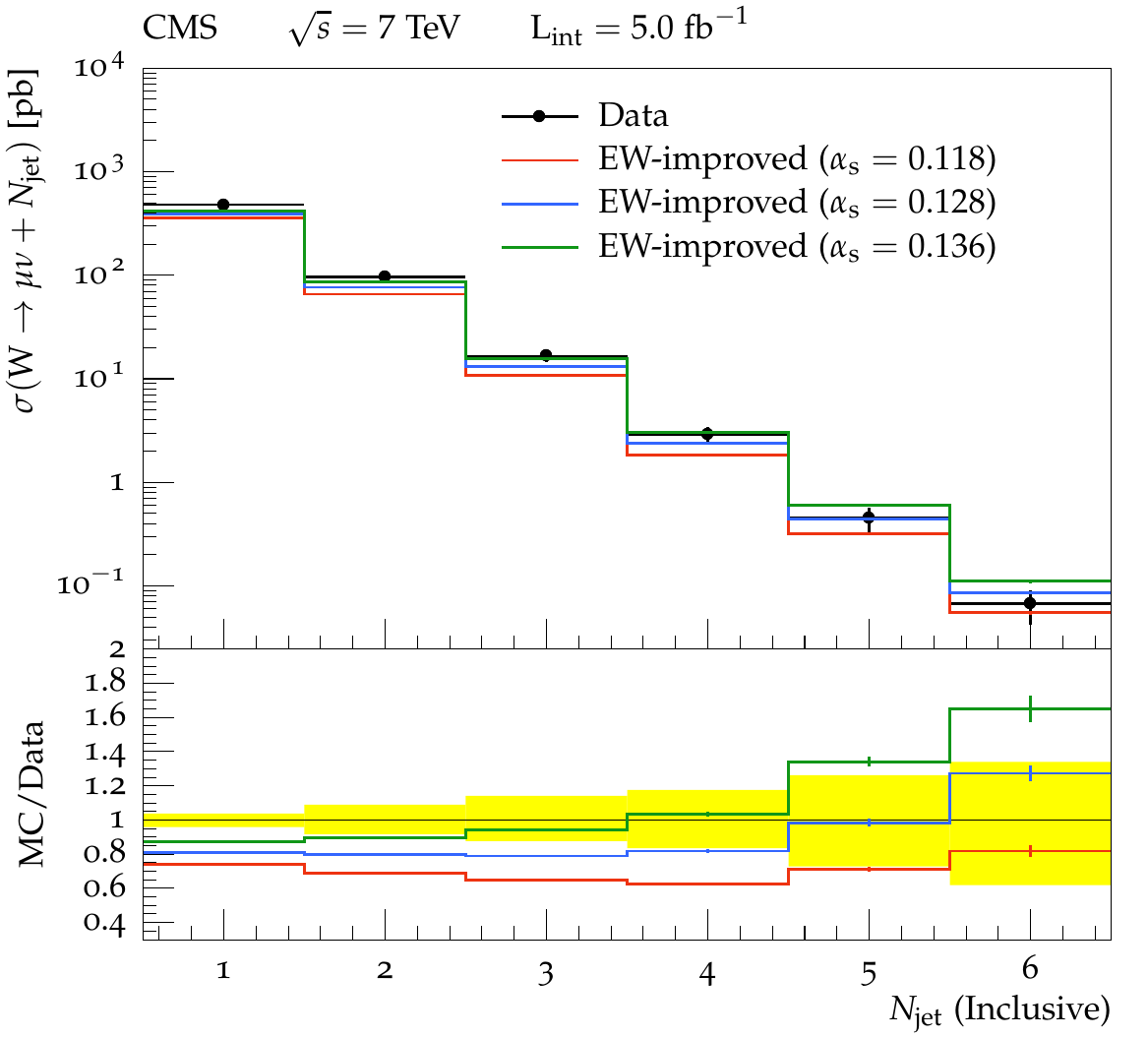}
  \caption{\Pythia predictions in comparison to ATLAS data~\cite{Aad:2014qxa}
    (left) and CMS data~\cite{Khachatryan:2014uva} (right)
    for W + jets as a function inclusive jet multiplicity. The yellow error
    band indicates the one sigma experimental uncertainty and the vertical
    line on the MC prediction is the statistical MC uncertainty. 
    ``$\as = 0.136$" stands for the values of the Monash 
    tune~\cite{Skands:2014pea}.
    \label{fig:aswJetMult}}
\end{figure}

The inclusive jet multiplicity in $\W$-boson events is well described by both the default and the
EW-improved merging (\figRef{fig:wJetMult}). The EW-improved model predicts a slightly lower 
cross section for large jet multiplicities, but given the large 
experimental uncertainty it is difficult to distinguish between the
models. Also, the value chosen for $\as(M_\Z)$ greatly influences the shape of
the distribution (\figRef{fig:aswJetMult}). The default value used in the Monash tune overshoots the
tail, whereas the PDG best fit value undershoots it. Choosing an in-between value
leads to a good agreement for all multiplicities. However, it should be noted
that $\as(M_Z)$ in the parton shower is tuned to jet-shapes (in $\e^+\e^-$ and 
hadron-hadron collisions) and just changing the value on a process-by-process 
means a significant loss of predictivity. An optimal solution would be a full 
retuning of the merged event generator to observables that have been measured for the 
purpose of tuning. This would result in a sensible best-fit value of $\as(M_Z)$ that
should be used for merged predictions of, say, the jet multiplicities. 
We refrain from providing a merged tune here, since distinguishing between the
``uncertainties" and the ``tunable parameters" of merged predictions is beyond the scope of
this study. All observables presented below are not
very sensitive to $\as(M_Z)$, up to overall normalisations. We thus
use $\as(M_\Z) = 0.118$ for all further studies.

\begin{figure}[tph]
  \centering
  \includegraphics[width=0.45\textwidth]{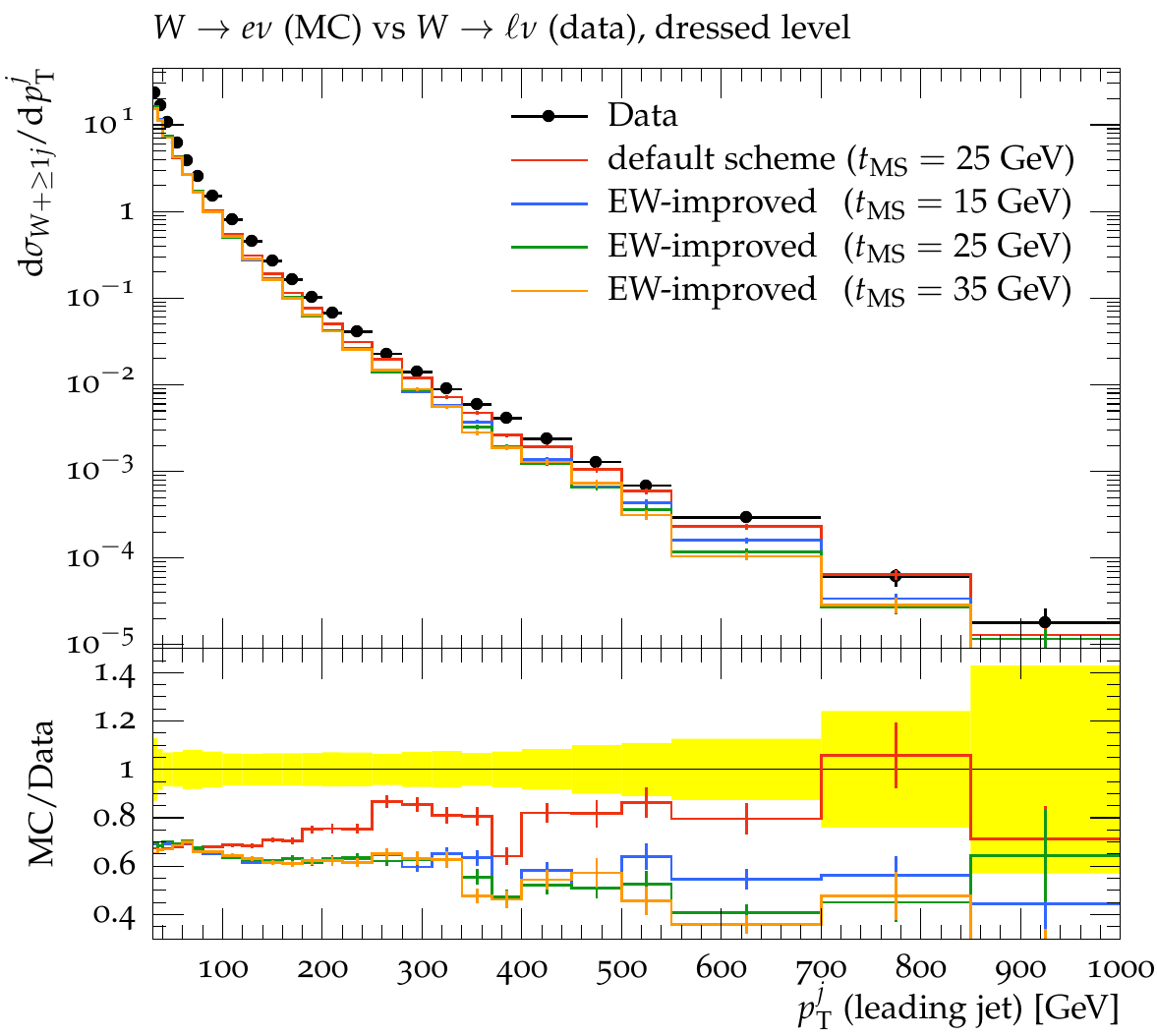}
  \includegraphics[width=0.45\textwidth]{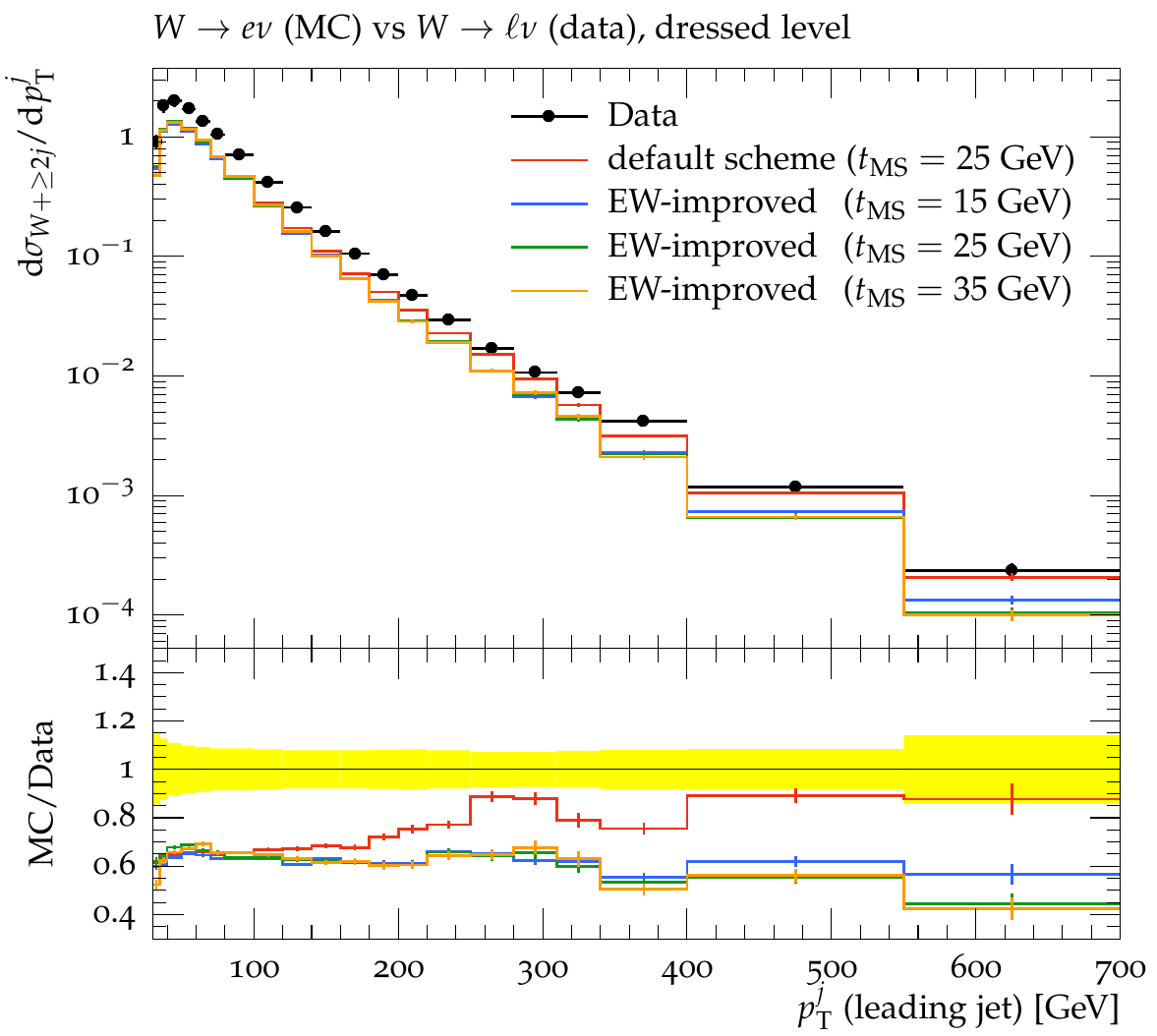}
  \caption{\Pythia predictions in comparison to ATLAS data~\cite{Aad:2014qxa}
    for W + jets as a function of the leading jet transverse momentum in inclusive
    $n-$jet events. The yellow error
    band indicates the one sigma experimental uncertainty and the vertical
    line on the MC prediction is the statistical MC uncertainty.
    \label{fig:wJetpT}}
\end{figure}

The $p_\perp$ distributions of individual jets (\figRef{fig:wJetpT}) provide a better test of
the default and EW-improved merging schemes. The fall-off observed in data is not
captured by the default model, whereas the EW-improved model describes the
shape of the data much better. This is the result of a more sensible scale 
setting, when the event is clustered to a
$2\rightarrow 2$ QCD process, and of the correct inclusion of the weak no-emission
probability. The default merging scheme had to compromise to determine the no-emission probability
for the unordered states. No such compromise is necessary now, since the
EW-improved scheme will instead naturally yield an underlying $2\rightarrow2$ QCD 
process and reweight accordingly. This clearly showcases that the ``merging
of mergings" scheme is favoured by data. 

\begin{figure}[tph]
  \centering
  \includegraphics[width=0.45\textwidth]{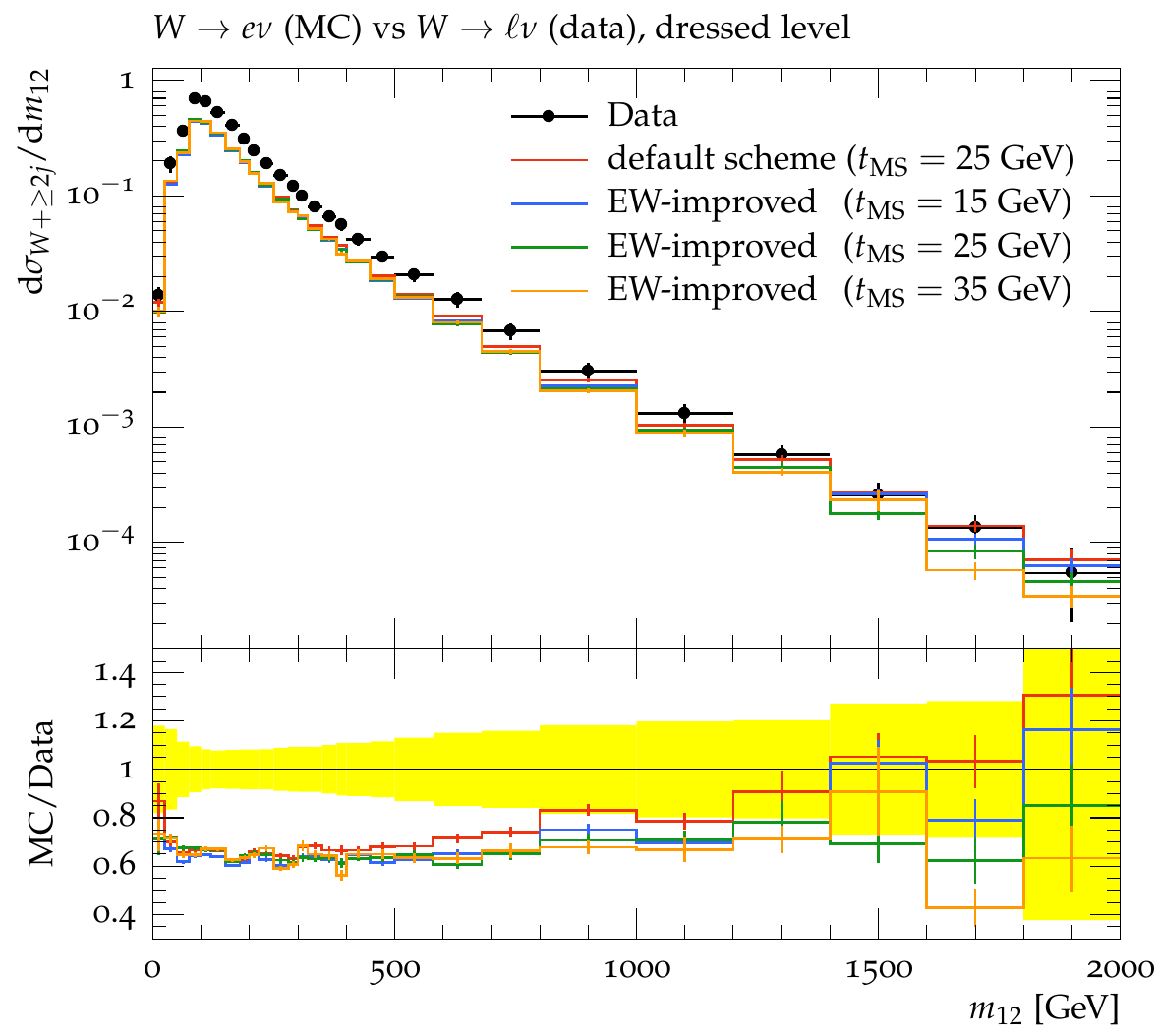}
  \includegraphics[width=0.45\textwidth]{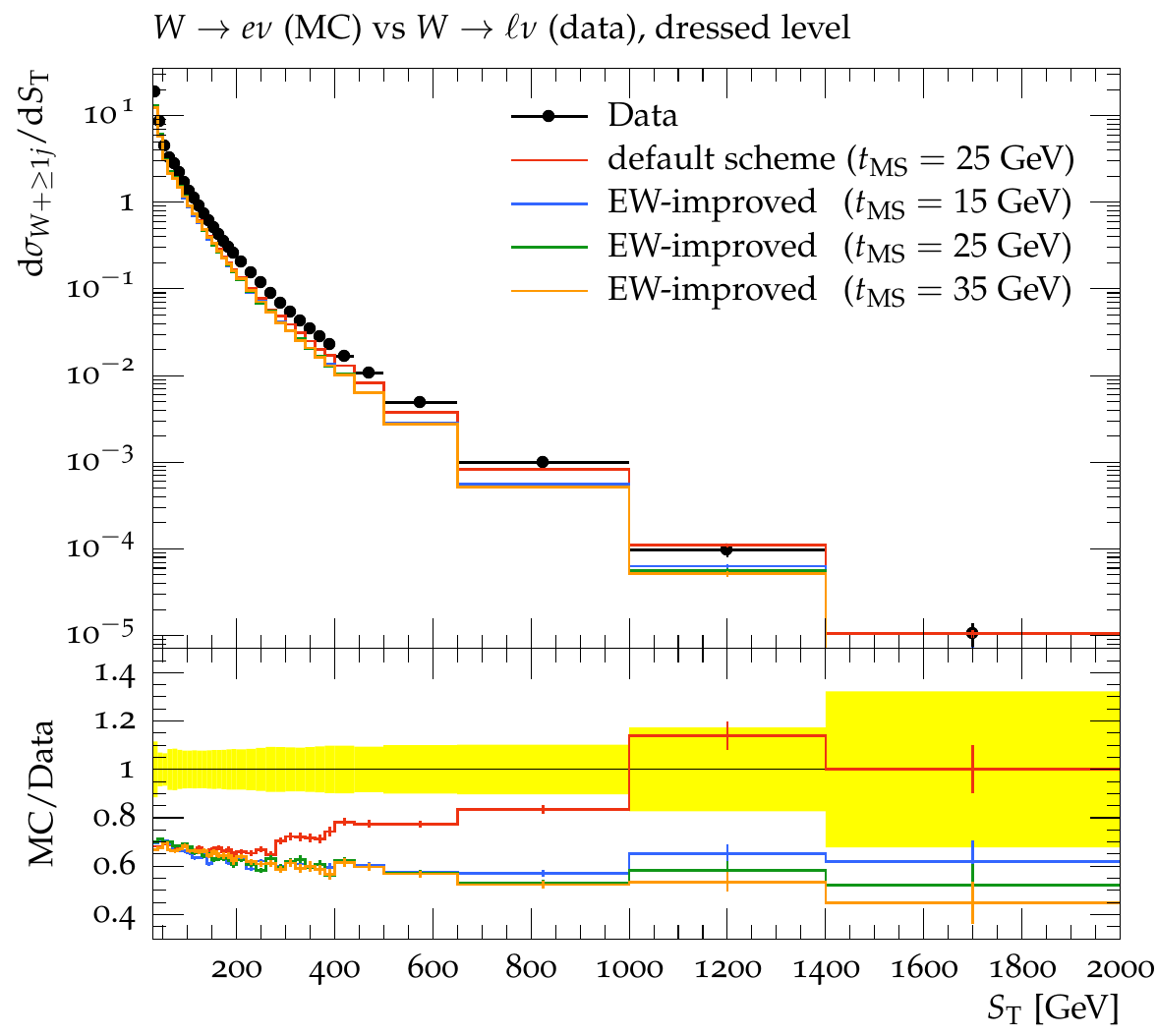}
  \caption{\Pythia predictions in comparison to ATLAS data~\cite{Aad:2014qxa}
    for W + jets as a function of $m_{12}$ and $S_T$. The
    yellow error band indicates the one sigma experimental uncertainty and the
    vertical line on the MC prediction is the statistical MC uncertainty.
    \label{fig:wJetInclusive}}
\end{figure}

More inclusive hardness-measures like the scalar $p_\perp$ sum of jets
$S_\textnormal{T}$ (\figRef{fig:wJetInclusive})
encourage the same conclusion. The effect is even more pronounced for these
observables. One of the observables that proved difficult to describe in the
original experimental study was the invariant mass between the two leading jets,
$m_{12}$. Again, the EW-improved merging scheme describes this observable well.

\begin{figure}[thp]
  \centering
  \includegraphics[width=0.45\textwidth]{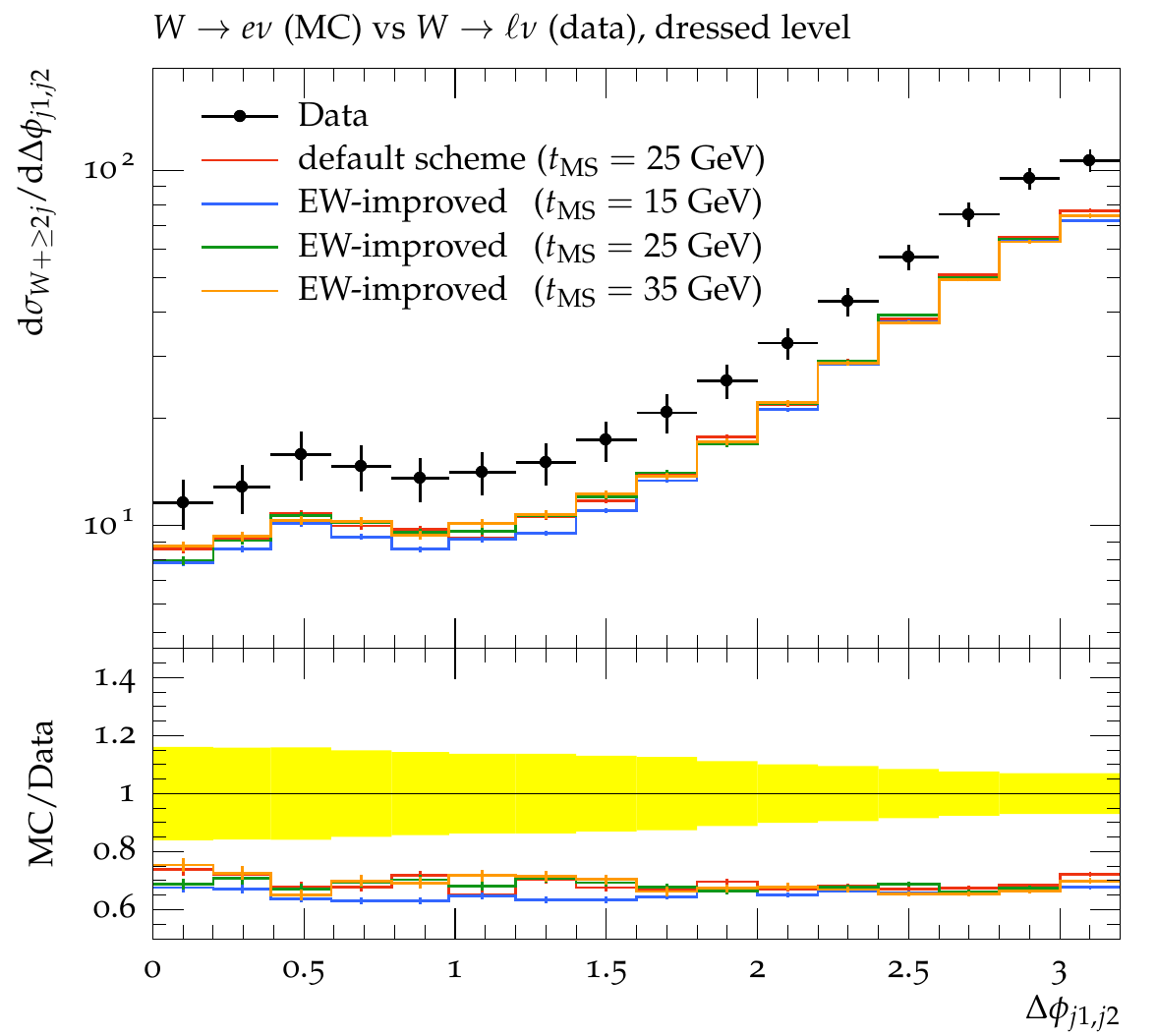}
  \includegraphics[width=0.45\textwidth]{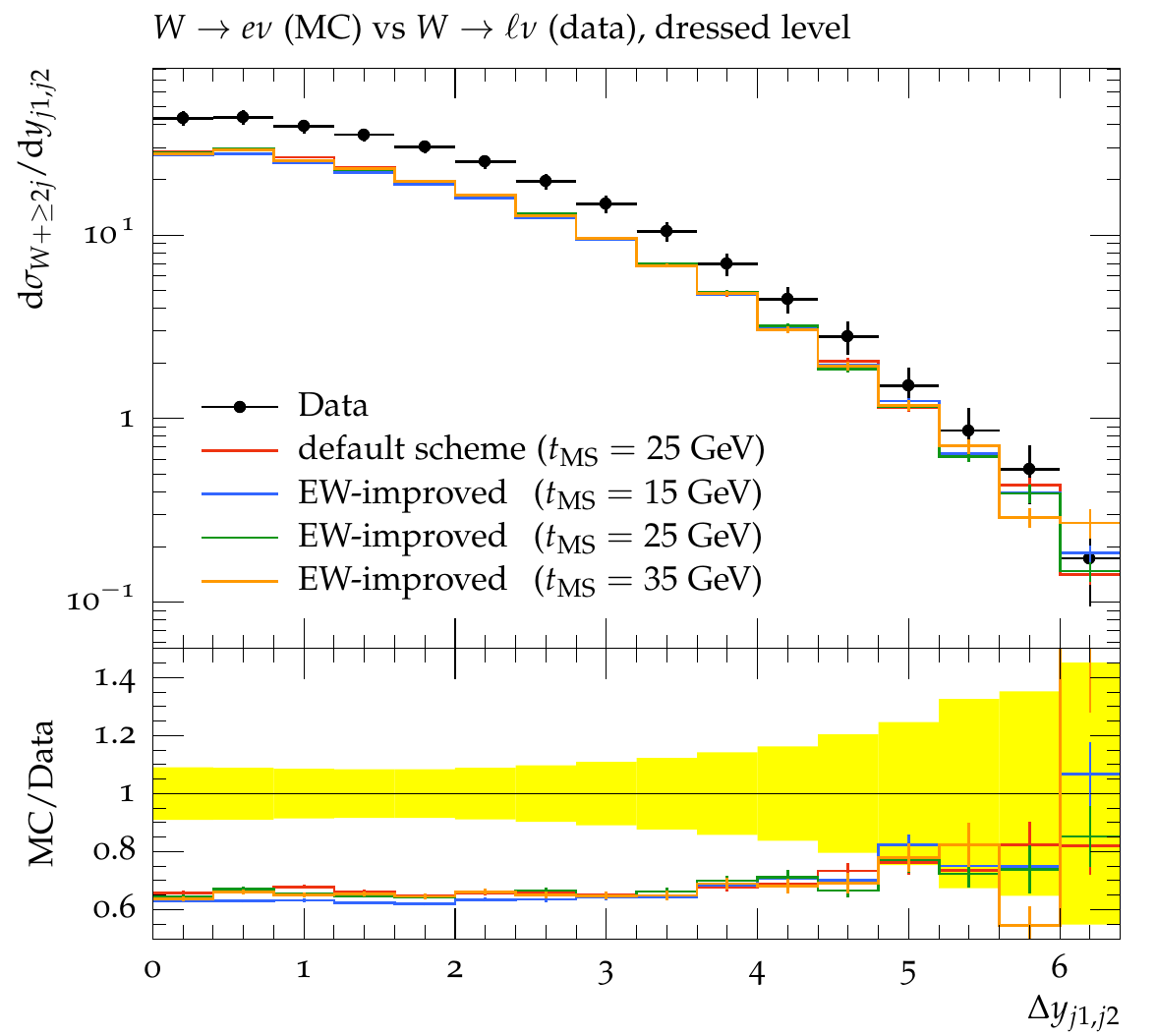}
  \caption{\Pythia predictions in comparison to ATLAS data~\cite{Aad:2014qxa}
    for W + jets as a function of $\Delta \phi_{12}$ and $\Delta y_{12}$. The yellow error band indicates the one sigma
    experimental uncertainty and the vertical line on the MC prediction is the
    statistical MC uncertainty. 
    \label{fig:wJetAngle}}
\end{figure}

Angular distributions are problematic for the weak parton showers. The
inclusion of merging is expected to improve this. This is exactly what is seen 
for $\Delta \phi_{12}$ and $\Delta y_{12}$ distributions
(\figRef{fig:wJetAngle}). Both the default and the EW-improved merging schemes provide
almost identical, and good, descriptions of the data.

\subsection{Predictions at 100 TeV}

When comparing with LHC data, we choose to highlight the importance of assigning
the correct underlying process, and disregarded other weak resummation effects to not
obscure the picture. In this section, we instead combine pure QCD multi-parton
states with W-boson + jets states. We therefore include W-bosons in the
merging scale cut: soft W-bosons will be produced by the shower,
while states containing hard W-bosons will be given by the fixed-order result. 

In order to assess the full effect of the merging
QCD+EW showers with multi-parton matrix elements, it is preferable to
consider 100 TeV pp collisions due to larger logarithmic enhancement with
increasing energy.
Observable that are commonly used to highlight weak resummation effects mostly
relate to exclusive dijet production.
However, in a combined resummation of QCD and EW logarithms, effects of
weak resummation will be completely dwarfed by all-order QCD. We will
therefore consider fully inclusive QCD and fully exclusive weak dijet 
production. Basically, whenever a weak boson is produced the event will not
enter the histograms. This should of course not be regarded as experimentally 
feasible, since a perfect W/Z tagging is doubtful. However, the setup can provide valuable insight
into the maximal size of effects related to the weak no-emission probability.
As event selection, we require at
least two jets with $p_\perp > 500$ GeV and the leading jet above $p_\perp >
1500$ GeV and no weak bosons.

\begin{figure}[thp]
  \centering
  \includegraphics[width=0.45\textwidth]{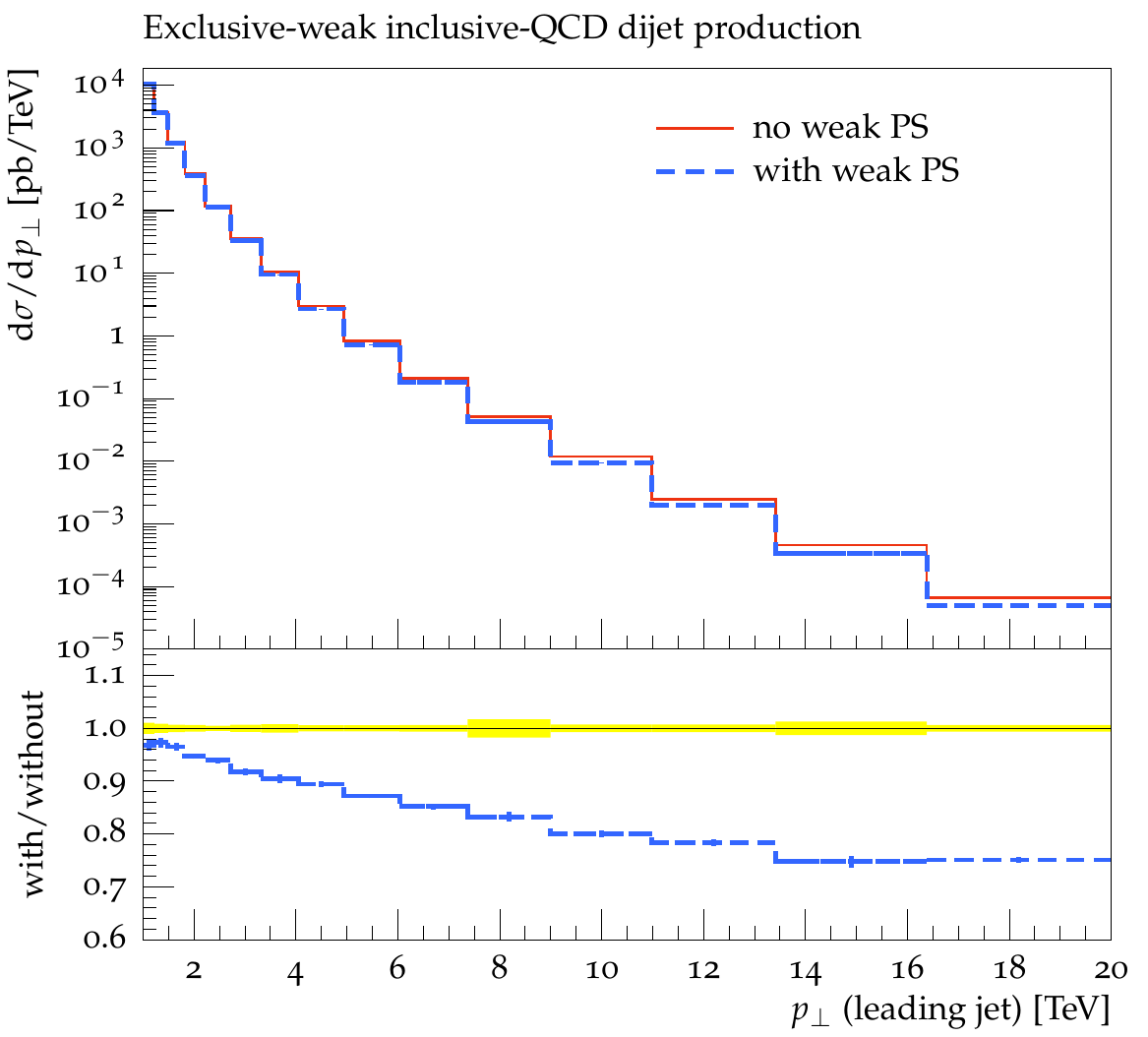}
  \includegraphics[width=0.45\textwidth]{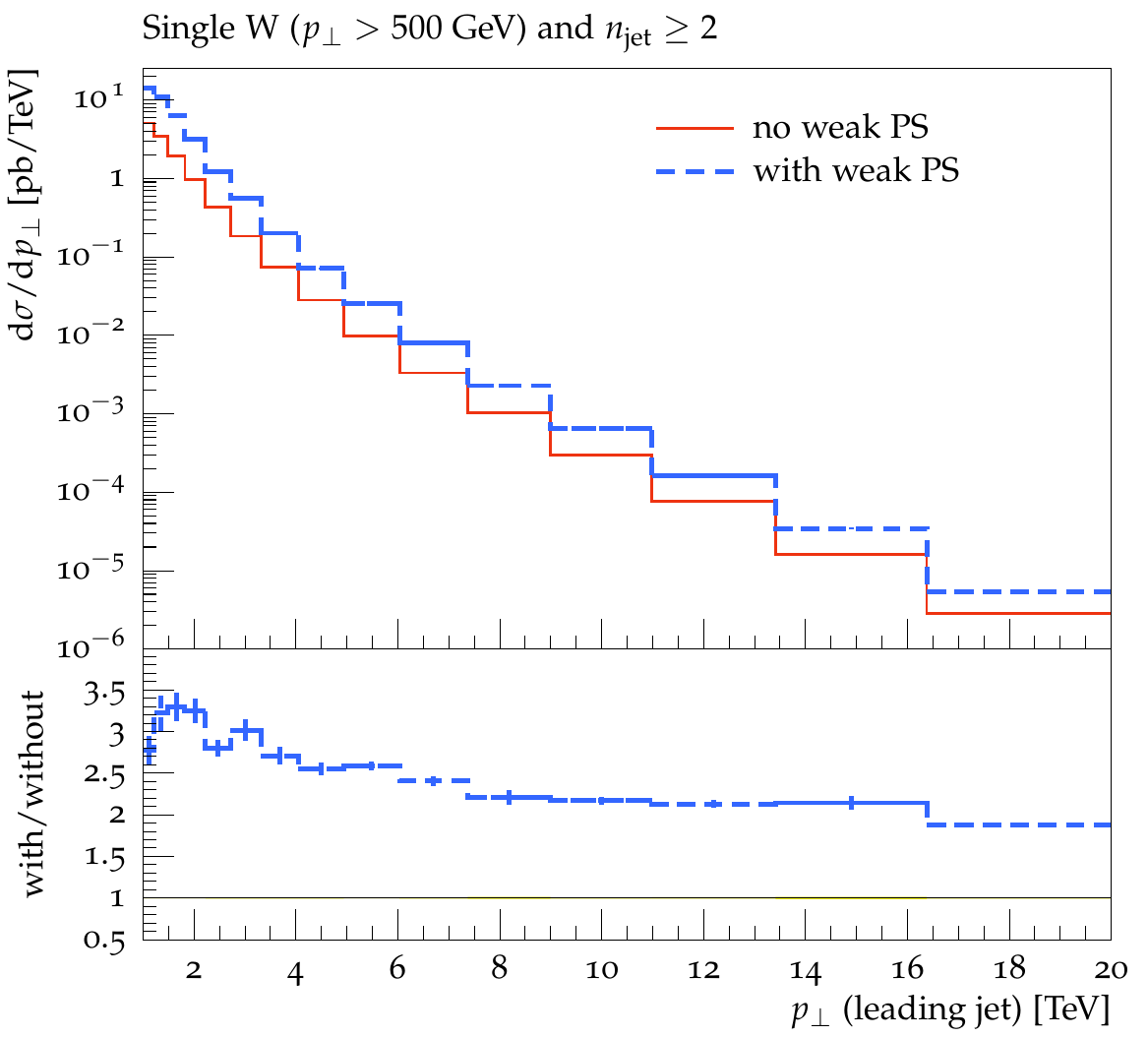}
  \caption{Predictions for 100 TeV respectively with and without including the
    weak PS for weak-exclusive dijet production (left) and weak-exclusive W +
    $\geq2$ jets (right). The yellow error band and the vertical lines indicate
    statistical MC uncertainty.
    \label{fig:100TeV}}
\end{figure}

The effects of the weak no-emission probability can seen 
in \figRef{fig:100TeV}, where we compare the result of including/not including
the weak PS when merging multi-jet with up to three outgoing partons.
The merging scales value is $\tms=500$ GeV. As expected the weak 
no-emission probability becomes more important for higher
$p_\perp$ scales and reaches roughly $25\%$ for a leading jet $p_\perp$ of 20 TeV. Even
at lower energies it might become important for high precision
measurements. This result is in agreement with the prediction from the
stand-alone weak PS~\cite{Christiansen:2014kba}.

A similar observable is the exclusive weak production of a $\W^\pm$ boson in association with at
least two jets (\figRef{fig:100TeV}). In addition to the multi-jet samples, this
simulation also requires W + $\leq 2$ jets samples. As such, this
simulation presents a fully inclusive merging of processes with vastly
different cross sections. The event selection applies the jet
selection outlined above, but additionally requires exactly one W with $p_\perp > $ 500 GeV, 
and no further weak bosons. The interpretation in terms of resummation 
effects is not as straightforward for this observable. The inclusion
of the weak PS both adds real radiation while simultaneously lowering
the cross section due to the inclusion of no-emission probabilities. However,
the real emission enhancement is overwhelming and leads to a factor of 2--3
enhancement of the cross section. This clearly shows the need for including
weak corrections, since the prediction from the ME alone is too low. If higher
jet multiplicities (e.g. W + 4 jet) are included, this effect is expected to
become milder. 

An earlier study~\cite{Mishra:2013una} of the electroweak corrections showed
significantly larger effects, reaching up to $\sim 80\,\%$ for lower jet
energies. This earlier study calculates the full EW NLO and compares the
differences between LO and NLO. The studies are not directly compatible, due
to different treatment of multiple effects, including: handling of photons,
competition between QCD and weak bosons, Bloch-Nordsieck violations~\cite{PhysRev.52.54}
and different analysis conditions. Further studies to get a better handle on
weak Sudakov effects would be of great interest.

\section{Conclusions}
\label{sec:conclusions}

We have presented a new consistent way of combining associated weak boson
radiation in hard dijet production with hard QCD radiation in Drell-Yan-type
events. It captures the strengths of both the merging technique and the weak
PS, while removing issues intrinsic to either. More specifically, we provide a
first matrix element merged prediction that consistently includes weak all-order 
effects. The combination of weak and QCD corrections leads to the concept of
a ``merging of mergings": processes with vastly different 
lowest-order cross sections are combined into a single consistent sample. We 
have addressed the problem of unordered states in this 
context, and a dynamical solution based on the dominance of certain scale 
hierarchies (i.e.\ evolution histories) in certain phase space regions has 
been presented. The novel prescription will 
be made available with the next release of \textsc{Pythia}~8.

The new merging scheme is compared to experimental data from ATLAS and
CMS. For all considered distributions the new EW-improved merging scheme does as least as
well as the old default merging. For a large fraction of the distributions, the EW-improved
scheme shows a significant improvement over the previous results. Especially for
high $S_\textnormal{T}$, where the EW-improved merging predicts a lower production rate by 
a more physical scale setting, by including weak no-emission probabilities, and
by identifying a $2\rightarrow2$ QCD scattering as underlying process.

The importance of the weak Sudakov for dijet production have been assessed in
the EW-improved merging scheme. The effects are shown to be about $25~\%$ at large
jet $p_\perp$ at a 100 TeV proton collider. Further studies comparing the
predicted corrections in the EW-improved merging scheme with NLO EW calculation would
be an interesting next step.

This study only includes the merging of $\W$ bosons and jets, but the Sudakov coming
from both $\W^\pm$ and $\Z^0$ bosons is accounted for. The implementation of $\Z^0$ in the
same merging framework is purely a technical, and is expected to simpler than
that of the $\W^\pm$.

In this study the weak merging scheme was only implemented for the CKKW-L
merging of leading-order matrix elements, as this relatively simple merging
method allows us to isolate and address generic problems without obscuring
the discussion by irrelevant details. A natural, and intriguing, 
next step is 
to extend the novel prescription to the UMEPS and UNLOPS schemes implemented
in \Pythia. Especially the latter would be of great interest, since it would
yield an event simulation that contains multiple NLO calculations for multiple
processes consistently combined with both QCD and EW resummation. The challenge
of such a generalisation is expected to be technical rather than conceptual.

\section*{Acknowledgments}

We thank Leif L{\"o}nnblad and Torbj{\"o}rn Sj{\"o}strand
for many enlightening discussions and for their comments on the manuscript.
SP gratefully acknowledges discussions about merging with Stefan H\"{o}che.
JRC is in part supported by Swedish Research Council, contract number
621-2013-4287, and in part by the MCnetITN FP7 Marie Curie Initial 
Training Network, contract PITN-GA-2012-315877. SP is supported by the US 
Department of Energy under contract DE–AC02–76SF00515.

\bibliographystyle{amsunsrt_modp}
\bibliography{journal}
\end{document}